\documentstyle[sprocl,epsf]{article}
\begin{document}
%\draft

\def\simge{\hspace*{0.2em}\raisebox{0.5ex}{$>$}
     \hspace{-0.8em}\raisebox{-0.3em}{$\sim$}\hspace*{0.2em}}
\def\simle{\hspace*{0.2em}\raisebox{0.5ex}{$<$}
     \hspace{-0.8em}\raisebox{-0.3em}{$\sim$}\hspace*{0.2em}}
\def\bra#1{{\langle#1\vert}}
\def\ket#1{{\vert#1\rangle}}
\def\coeff#1#2{{\scriptstyle{#1\over #2}}}
\def\undertext#1{{$\underline{\hbox{#1}}$}}
\def\hcal#1{{\hbox{\cal #1}}}
\def\sst#1{{\scriptscriptstyle #1}}
\def\eexp#1{{\hbox{e}^{#1}}}
\def\rbra#1{{\langle #1 \vert\!\vert}}
\def\rket#1{{\vert\!\vert #1\rangle}}
\def\lsim{{ <\atop\sim}}
\def\gsim{{ >\atop\sim}}
\def\nubar{{\bar\nu}}
\def\psibar{{\bar\psi}}
\def\Gmu{{G_\mu}}
\def\alr{{A_\sst{LR}}}
\def\wpv{{W^\sst{PV}}}
\def\evec{{\vec e}}
\def\notq{{\not\! q}}
\def\notl{{\not\! \ell}}
\def\notk{{\not\! k}}
\def\notp{{\not\! p}}
\def\notpp{{\not\! p'}}
\def\notder{{\not\! \partial}}
\def\notcder{{\not\!\! D}}
\def\notA{{\not\!\! A}}
\def\notv{{\not\!\! v}}
\def\Jem{{J_\mu^{em}}}
\def\Jana{{J_{\mu 5}^{anapole}}}
\def\nue{{\nu_e}}
\def\mn{{m_\sst{N}}}
\def\mns{{m^2_\sst{N}}}
\def\me{{m_e}}
\def\mes{{m^2_e}}
\def\mq{{m_q}}
\def\mqs{{m_q^2}}
\def\mz{{M_\sst{Z}}}
\def\mzs{{M^2_\sst{Z}}}
\def\ubar{{\bar u}}
\def\dbar{{\bar d}}
\def\sbar{{\bar s}}
\def\qbar{{\bar q}}
\def\sstw{{\sin^2\theta_\sst{W}}}
\def\gv{{g_\sst{V}}}
\def\ga{{g_\sst{A}}}
\def\pv{{\vec p}}
\def\pvs{{{\vec p}^{\>2}}}
\def\ppv{{{\vec p}^{\>\prime}}}
\def\ppvs{{{\vec p}^{\>\prime\>2}}}
\def\qv{{\vec q}}
\def\qvs{{{\vec q}^{\>2}}}
\def\xv{{\vec x}}
\def\xpv{{{\vec x}^{\>\prime}}}
\def\yv{{\vec y}}
\def\tauv{{\vec\tau}}
\def\sigv{{\vec\sigma}}
\def\sst#1{{\scriptscriptstyle #1}}
\def\gpnn{{g_{\sst{NN}\pi}}}
\def\grnn{{g_{\sst{NN}\rho}}}
\def\gnnm{{g_\sst{NNM}}}
\def\hnnm{{h_\sst{NNM}}}

\def\xivz{{\xi_\sst{V}^{(0)}}}
\def\xivt{{\xi_\sst{V}^{(3)}}}
\def\xive{{\xi_\sst{V}^{(8)}}}
\def\xiaz{{\xi_\sst{A}^{(0)}}}
\def\xiat{{\xi_\sst{A}^{(3)}}}
\def\xiae{{\xi_\sst{A}^{(8)}}}
\def\xivtez{{\xi_\sst{V}^{T=0}}}
\def\xivteo{{\xi_\sst{V}^{T=1}}}
\def\xiatez{{\xi_\sst{A}^{T=0}}}
\def\xiateo{{\xi_\sst{A}^{T=1}}}
\def\xiva{{\xi_\sst{V,A}}}

\def\rvz{{R_\sst{V}^{(0)}}}
\def\rvt{{R_\sst{V}^{(3)}}}
\def\rve{{R_\sst{V}^{(8)}}}
\def\raz{{R_\sst{A}^{(0)}}}
\def\rat{{R_\sst{A}^{(3)}}}
\def\rae{{R_\sst{A}^{(8)}}}
\def\rvtez{{R_\sst{V}^{T=0}}}
\def\rvteo{{R_\sst{V}^{T=1}}}
\def\ratez{{R_\sst{A}^{T=0}}}
\def\rateo{{R_\sst{A}^{T=1}}}

\def\mro{{m_\rho}}
\def\mks{{m_\sst{K}^2}}
\def\mpi{{m_\pi}}
\def\mpis{{m_\pi^2}}
\def\mom{{m_\omega}}
\def\mphi{{m_\phi}}
\def\Qhat{{\hat Q}}

\def\FOS{{F_1^{(s)}}}
\def\FTS{{F_2^{(s)}}}
\def\GAS{{G_\sst{A}^{(s)}}}
\def\GES{{G_\sst{E}^{(s)}}}
\def\GMS{{G_\sst{M}^{(s)}}}
\def\GATEZ{{G_\sst{A}^{\sst{T}=0}}}
\def\GATEO{{G_\sst{A}^{\sst{T}=1}}}
\def\mdax{{M_\sst{A}}}
\def\mustr{{\mu_s}}
\def\rsstr{{r^2_s}}
\def\rhostr{{\rho_s}}
\def\GEG{{G_\sst{E}^\gamma}}
\def\GEZ{{G_\sst{E}^\sst{Z}}}
\def\GMG{{G_\sst{M}^\gamma}}
\def\GMZ{{G_\sst{M}^\sst{Z}}}
\def\GEn{{G_\sst{E}^n}}
\def\GEp{{G_\sst{E}^p}}
\def\GMn{{G_\sst{M}^n}}
\def\GMp{{G_\sst{M}^p}}
\def\GAp{{G_\sst{A}^p}}
\def\GAn{{G_\sst{A}^n}}
\def\GA{{G_\sst{A}}}
\def\GETEZ{{G_\sst{E}^{\sst{T}=0}}}
\def\GETEO{{G_\sst{E}^{\sst{T}=1}}}
\def\GMTEZ{{G_\sst{M}^{\sst{T}=0}}}
\def\GMTEO{{G_\sst{M}^{\sst{T}=1}}}
\def\lamd{{\lambda_\sst{D}^\sst{V}}}
\def\lamn{{\lambda_n}}
\def\lams{{\lambda_\sst{E}^{(s)}}}
\def\bvz{{\beta_\sst{V}^0}}
\def\bvo{{\beta_\sst{V}^1}}
\def\Gdip{{G_\sst{D}^\sst{V}}}
\def\GdipA{{G_\sst{D}^\sst{A}}}
\def\fks{{F_\sst{K}^{(s)}}}
\def\FIS{{F_i^{(s)}}}
\def\fpi{{F_\pi}}
\def\fk{{F_\sst{K}}}

\def\RAp{{R_\sst{A}^p}}
\def\RAn{{R_\sst{A}^n}}
\def\RVp{{R_\sst{V}^p}}
\def\RVn{{R_\sst{V}^n}}
\def\rva{{R_\sst{V,A}}}
\def\xbb{{x_B}}

\def\mlq{{M_\sst{LQ}}}
\def\mlqs{{M_\sst{LQ}^2}}
\def\lscal{{\lambda_\sst{S}}}
\def\lvect{{\lambda_\sst{V}}}

\def\PR#1{{{\em   Phys. Rev.} {\bf #1} }}
\def\PRC#1{{{\em   Phys. Rev.} {\bf C#1} }}
\def\PRD#1{{{\em   Phys. Rev.} {\bf D#1} }}
\def\PRL#1{{{\em   Phys. Rev. Lett.} {\bf #1} }}
\def\NPA#1{{{\em   Nucl. Phys.} {\bf A#1} }}
\def\NPB#1{{{\em   Nucl. Phys.} {\bf B#1} }}
\def\AoP#1{{{\em   Ann. of Phys.} {\bf #1} }}
\def\PRp#1{{{\em   Phys. Reports} {\bf #1} }}
\def\PLB#1{{{\em   Phys. Lett.} {\bf B#1} }}
\def\ZPA#1{{{\em   Z. f\"ur Phys.} {\bf A#1} }}
\def\ZPC#1{{{\em   Z. f\"ur Phys.} {\bf C#1} }}
\def\etal{{{\em   et al.}}}

\def\delalr{{{delta\alr\over\alr}}}
\def\pbar{{\bar{p}}}
\def\lamchi{{\Lambda_\chi}}

\def\qw0{{Q_\sst{W}^0}}
\def\qwp{{Q_\sst{W}^P}}
\def\qwn{{Q_\sst{W}^N}}
\def\qwe{{Q_\sst{W}^e}}
\def\qem{{Q_\sst{EM}}}

\def\gae{{g_\sst{A}^e}}
\def\gve{{g_\sst{V}^e}}
\def\gvf{{g_\sst{V}^f}}
\def\gaf{{g_\sst{A}^f}}
\def\gvu{{g_\sst{V}^u}}
\def\gau{{g_\sst{A}^u}}
\def\gvd{{g_\sst{V}^d}}
\def\gad{{g_\sst{A}^d}}

\def\gvftil{{\tilde g_\sst{V}^f}}
\def\gaftil{{\tilde g_\sst{A}^f}}
\def\gvetil{{\tilde g_\sst{V}^e}}
\def\gaetil{{\tilde g_\sst{A}^e}}
\def\gvqtil{{\tilde g_\sst{V}^e}}
\def\gaqtil{{\tilde g_\sst{A}^e}}
\def\gvutil{{\tilde g_\sst{V}^e}}
\def\gautil{{\tilde g_\sst{A}^e}}
\def\gvdtil{{\tilde g_\sst{V}^e}}
\def\gadtil{{\tilde g_\sst{A}^e}}

\def\hvf{{h_\sst{V}^f}}
\def\hvu{{h_\sst{V}^u}}
\def\hvd{{h_\sst{V}^d}}
\def\hve{{h_\sst{V}^e}}
\def\hvq{{h_\sst{V}^q}}

\def\delp{{\delta_P}}
\def\delzp{{\delta_{00}}}
\def\deld{{\delta_\Delta}}
\def\dele{{\delta_e}}

\def\apv{{A_\sst{PV}}}
\def\apvnsid{{A_\sst{PV}^\sst{NSID}}}
\def\apvnsd{{A_\sst{PV}^\sst{NSD}}}
\def\qpv{{Q_\sst{W}}}

\title{\hfill{\normalsize INT \#DOE/ER/40561-37-INT98}\\[2.5cm]
SENSITIVITY OF LOW-ENERGY PARITY-VIOLATION TO NEW PHYSICS}

\author{M.J. RAMSEY-MUSOLF$^{a,b}$
%\thanks{National Science Foundation Young Investigator}\\[0.3cm]
}
\address{
$^a$ Department of Physics, University of Connecticut, 
Storrs, CT 06269 USA\\
$^b$ Institute for Nuclear Theory, 
University of Washington, Seattle, WA 98195 USA \\
email: mjrm@phys.uconn.edu}

%\date

\maketitle

\abstracts{
I review the new physics sensitivity of low-energy parity-violating (PV) 
observables. I
concentrate on signatures of new tree-level physics in atomic PV with
a single isotope, ratios of atomic PV observables, and PV electron scattering. 
In addition
to comparing the new physics sensitivities of these observables with those of 
high-energy
colliders, I also discuss the theoretical issues involved in the extraction of 
new physics
limits from low-energy PV observables.} 

%\end{abstract}

%\pacs{}

%\narrowtext

\section{Introduction}
\label{sec:intro}

From its discovery in the $\beta$-decay of $^{60}$Co by Wu {\em et al.} in 1957, 
parity-violation
(PV) in nuclear and atomic processes has played a central role in elucidating 
the structure of
the electroweak interaction. By now, our gauge theory of that interaction -- the 
SU(2$)_L
\times$U(1$)_Y$ Standard Model -- has been tested in a wide variety of 
processes, ranging in
energies from the eV scale to the 100 GeV scale. The agreement between 
experiment and the
predictions of the Standard Model (SM) is impressive. In nearly all cases, there 
is accord at the
0.1\% level or better. A striking illustration is the discovery of the top 
quark, whose measured
mass falls within a rather narrow range predicted from global analysis of 
electroweak observables
at the level of one-loop SM radiative corrections~\cite{Sch95}. There are, 
however, a few
exceptions to this pattern of agreement, such as the value of the 
Cabbibo-Kobayashi-Maskawa
matrix element 
$V_{ud}$ determined from nuclear $\beta$-decay. Analyses of superallowed 
$\beta$-decays imply a
value for $|V_{ud}|$ differing from the SM unitarity requirement by nearly two 
standard
deviations~\cite{Tow95}. Whether this discrepancy is due to a difficiency in the 
SM or an unknown
systematic in $\beta$-decay analyses remains an open question. Despite this -- 
and a few other --
apparent disagreements, the SM works incredibly well for most of what is 
observed experimentally.

These days, experiments at high-energy colliders (LEP, Tevatron, HERA) are 
engaged in searches
for physics \lq\lq beyond" the SM. The motivation for seeking such \lq\lq new" 
physics is that
the Standard Model is just that -- a model. As well as it works in describing 
and predicting
electroweak processes, it also leaves several questions unanswered. It requires 
as input, for
example, 17 independent, experimentally-determined parameters, but doesn't tell 
us why these
parameters take on their measured values. Why, for example, is $m_e<<m_\mu << 
m_\tau$ (hierarchy
problem) ? Similarly, the SM assumes elementary fermions and bosons to have no 
size or structure,
but does not tell us why this assumption must be true. It incorporates PV as 
observed, but does
not explain {\em why} parity is violated. When the electroweak and strong 
sectors of the SM are
taken together, the couplings associated with each interaction do not meet at a 
common point when
run to high scales $\mu$. This absence of unification is theoretically 
un-satisfying at best. Such
questions and \lq\lq loose ends" suggest that teh SM is really a low-energy 
($<<$ weak scale)
effective theory of some more general framework -- one which contains, in 
principle, physics we
have not searched for with sufficient intention. Hence, there exists 
considerable interest in
searching for this new physics.

What I hope to show in this chapter is how PV on a table top (atoms) and in 
low-energy colliders
(electron scattering) is a powerful probe of possible new physics -- and one 
which generally
complements high-energy collider searches. The reason for PV's continued 
relevance is the high
precision with which measurements can be performed. In this respect, the 
benchmark
experiment is the one in atomic parity-violation (APV) performed on cesium by 
the Boulder group
\cite{Woo97}. The experimental error in that measurement is an impressive 
0.35\%. 
 Unfortunately,
the {\em interpretation} of the cesium result requires input from atomic theory, 
for which the
present uncertainty is $\sim 1.2\%$. Future progress in using APV to probe for 
new physics will
require signficant reduction in that uncertainty. Alternatively, several groups 
are pursuing
measurements of {\em ratios} of APV observables for different atoms along the 
isotope chain
\cite{Bud98}. The atomic theory-dependence of these \lq\lq isotope ratios", 
${\cal R}$, largely
cancels out, leaving an atomic theory-free probe of new physics. The isotope 
ratios, however,
display a significant dependence on the neutron number density, $\rho_n(r)$, of 
atomic nuclei.
The neutron densities are not sufficiently well-determined experimentally, so 
one must rely on
{\em nuclear} theory to compute them. Whether or not the nuclear theory 
uncertainty is
sufficiently small to interpret
${\cal R}$ in terms of new physics remains an open question. As I will 
illustrate toward the end
of the chapter, PV $e$-$e$ and $e$-hadron scattering offers the theoretically 
cleanest probe of
new physics. It remains to be see whether $\alr$, the \lq\lq left-right" 
asymmetry in PV electron
scattering (PVES), can be measured with the precision needed to make it 
competitive with APV. 
The recent successes of the Jefferson Laboratory PV experiment suggests that the
possibilities are promising. 

In the remainder of the chapter, my discussion of these points is organized as 
follows.
In Section II, I introduce some formal definitions and conventions needed for 
the subsequent
discussion. In Section III, I consider a variety of low-energy PV observables 
and compare their
generic sensitivities to new physics. In Section IV, I illustrate these 
sensitivities with
different model scenarios for physics beyond the Standard Model. Section V 
contains a discussion
of theoretical uncertainties which arise in the interpretation of the PV 
observables. Section VI
summarizes my principal conclusions. Much of what I discuss here is also treated 
more fully in
Refs. \cite{Mus94,MRM98}, and I refer the reader to those papers for additional 
details.

\section{Formalism}
\label{sec:formal}

For purposes of probing new physics with PV, the quantity of interest is the 
weak charge,
$\qpv$. It is the weak neutral current analog of the electromagnetic charge, 
$\qem$. The weak
charge appears in the amplitude associated with Fig. 1a, where an electron 
interacts with another fermion by exchanging a $Z^0$ boson.
\begin{figure}
\epsfbox{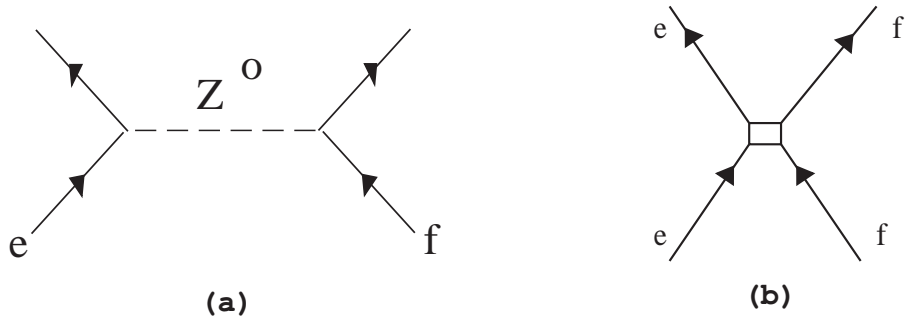}
\caption{\label{Fig1} (a) Parity-violating electron-fermion amplitude generated 
by
$Z^0$-exchange. (b) Effective, four-fermion electron-fermion PV interaction.}
\end{figure}
If $V(f)$ and $A(f)$ denote the weak 
neutral vector and
axial vector currents, respectively, of an elementary fermion $f$, then the PV 
part of the
amplitude in Fig. 1a depends on either $A(e)\times V(f)$ or $V(e)\times A(f)$. 
The former is 
characterized by $Q_\sst{W}^f$, the coupling of $V(f)$ to the $Z^0$ boson. When 
an electron
interacts weakly with a system composed of several elementary fermions, such as 
an atomic
nucleus, the weak charge of that system, $\qpv$, is just the sum of the 
$Q_\sst{W}^f$ of its
constituents. In this respect, $\qpv$ behaves just like $\qem$. However, because 
electroweak
symmetry is broken below the weak scale ($\simle 250$ GeV), leaving only 
electromagnetic gauge
invariance as a good symmetry, $\qpv$ is {\em not conserved}. Its value can be 
predicted by the
SM, but that value may be altered by the presence of new physics. The EM charge, 
on the other
hand, is protected from such new physics modifications by EM gauge invariance. 

The new physics modification of $\qpv$ can arise in basically two ways: by 
physics which
modifies the propagation of the $Z^0$ from the electron to the other fermion, 
and by the exchange
of possible new, heavy particles between $e$ and $f$. Modifications of the 
former type are called
\lq\lq oblique", whereas the latter induce so-called \lq\lq direct" new physics 
corrections to
$\qpv$. In general, low-energy PV constitutes a much more powerful probe -- 
relative to other
electroweak processes -- of direct new physics than of oblique new physics 
\cite{Ros98}.
Consequently, in what follows, I will not treat oblique corrections and 
concentrate solely on new
direct interactions. To that end, I remind the reader that at low-energies, one 
has
$|q^2|<<\mzs$, where $q_\mu$ is the momentum-transfer from the electron to the 
fermion (or
nucleus) in Fig. 1a. Consequently, in a low-energy experiment, the process of 
Fig. 1a looks like
it arises from a four fermion contact interaction, whose strength is 
proportional to $1/\mzs$
(Fig. 1b). Similarly, any new direct interactions, characterized by a heavy mass 
scale $\Lambda$
({\em e.g.}, the mass of another exchanged particle) will also look like a four 
fermion
interaction whose strength is proportional to $1/\Lambda^2$. Consequently, we 
may write 
\begin{equation}
\qpv=\qw0+\Delta\qpv \ \ \ .
\end{equation}
Here, $\qw0$ gives the contribution in the Standard Model while 
$\Delta\qpv$ indicates possible contributions from new interactions. We consider
$\qpv$ to be generated by the low-energy effective Lagrangian
\begin{equation}
\label{form:qwa}
{\cal L}={\cal L}^\sst{PV}_\sst{S.M.}+{\cal L}^\sst{PV}_\sst{NEW}\ \ \ ,
\end{equation}
where
\begin{eqnarray}
\label{form:lsm}
{\cal L}^\sst{PV}_\sst{S.M.}&=&{G_F\over 2\sqrt{2}}\gae{\bar e}\gamma_\mu
	\gamma_5 e\sum_f \gvf{\bar f}\gamma^\mu f \\
\label{form:lnew}
{\cal L}^\sst{PV}_\sst{NEW}&=&{4\pi\kappa^2\over\Lambda^2} {\bar e}
	\gamma_\mu\gamma_5 e\sum_f\hvf{\bar f}
	\gamma^\mu f\ \ \ .
\end{eqnarray}
Here $\gvf\equiv Q_\sst{W}^f =2T_3^f-4Q_f\sstw$ and $\gaf=-2T_3^f$ are the tree 
level
vector and axial vector
fermion-$Z^0$ couplings in the SM, with $Q_f$ being the fermion's EM charge and
$T_3^f$ its weak isospin. The coupling $\hvf$ characterizes the 
interaction of the electron axial vector current with the vector
current of fermion $f$ for a given extension of the Standard Model. As above, 
$\Lambda$ is the mass scale associated with the new physics, while
$\kappa$ sets the coupling strength. Generally speaking, strongly interacting
theories take $\kappa^2\sim 1$ while for weakly interacting extensions of the
Standard Model one has $\kappa^2\sim\alpha$. For scenarios in which 
the interaction of Eq. (\ref{form:lnew}) is generated by the exchange of a new
heavy particle between the electron and fermion, the constant
$\hvf=\gaetil\gvftil$, where $\gaetil$ ($\gvftil$) are the heavy
particle axial vector (vector) coupling to the electron (fermion). 

For simplicity, I do not consider contributions to $\Delta\qpv$ arising
from new scalar-pseudoscalar or tensor-pseudotensor interactions.
I also do not consider $V(e)\times A(f)$ interactions, as they do not
contribute to $\qpv$. Although the Standard Model $V(e)\times A(f)$
interaction is suppressed due to the small value of $\gae=-1+4\sstw$,
resulting in an enhanced sensitivity to new physics of this type, one
is at present not able to extract the $V(e)\times A(f)$ amplitudes
from PV observables
with the level of precision attainable for $\qpv$ (see, {\em e.g.}, Ref. 
\cite{Muk98}). Moreover,
the hadronic axial vector current is not protected by current conservation from 
hadronic
effects which may cloud the interpretation of the hadronic axial vector
amplitude in terms of new physics \cite{Mus90}. For the interaction of an 
electron with
a heavy nucleus, such as cesium, the $V(e)\times A(f)$ interaction is 
interesting
from another standpoint. As I discuss briefly in the next section, this term 
receives a sizeable
contribution from PV quark-quark interactions {\em inside} the nucleus. These PV 
hadronic
effects are discussed in greater detail elsewhere in this book. 

It is straightforward to write down the corrections to the weak charge of
a given system arising from ${\cal L}^\sst{PV}_\sst{NEW}$. Specifically,
consider the nucleon and electron:
\begin{eqnarray}
\label{form:dqwp}
\Delta\qwp&=&\zeta(2\hvu+\hvd)\\
\label{form:dqwn}
\Delta\qwn&=&\zeta(\hvu+2\hvd)\\
\label{form:dqwe}
\Delta\qwe&=&\zeta\hve\ \ \ ,
\end{eqnarray}
where
\begin{equation}
\label{form:scale}
\zeta = {8\sqrt{2}\pi\kappa^2\over\Lambda^2 G_F}\ \ \ .
\end{equation}

To obtain a feel for the sensitivity of $\qpv$ to the mass scale $\Lambda$, we 
may consider a
generic scenario in which the couplings
$\gvf$ and
$\hvf$ entering
$\qpv$ and
$\Delta\qpv$ are of the same order of magnitude. In this case, the fractional 
correction
induced by new physics is
\begin{equation}
\label{form:ratio}
{\Delta\qpv\over\qw0}={8\sqrt{2}\pi\over\Lambda^2 G_F}\ \ \ .
\end{equation}
If a determination of $\qpv$ is made with one percent uncertainty, then the 
ratio in Eq.
(\ref{form:ratio}) must be $\leq 0.01$. Re-arranging the inequality leads to a 
lower bound on
$\Lambda$:

\begin{equation}
\label{form:lam}
\Lambda\geq\left[{8\sqrt{2}\pi\kappa^2\over 0.01 G}\right]^{1/2}
\approx 20\kappa \ \ {\hbox{TeV}}\ \ \ .
\end{equation}
In short, determinations of $\qpv$ at the one percent or better level
probe new physics at the TeV scale for weakly interacting theories and
the ten TeV scale for new strong interactions.

\section{Observables}
\label{sec:observe}

Before analyzing the possible impact different new physics scenarios may have on
low-energy PV, it is instructive to consider the new physics sensitivities of  
different
observables in general terms. Specifically, I will consider a
general atomic PV observable for a single isotope, $\apv(N)$; ratios 
involving $\apv$ for different isotopes, ${\cal R}$; and the left-right
asymmetry for scattering polarized electrons from a given target, 
$\alr$. Of these, the simplest is the atomic PV observable for a single
isotope. In general, $\apv(N)$ contains a term which varies with the nuclear
spin, $\apvnsd(N)$, and a term independent of the nuclear spin, $\apvnsid(N)$. 
These two terms arise from the way the PV electron-nucleus interaction causes 
atomic
states of definite parity to mix. Because the $Z^0$ exchange interaction is a 
contact
interaction in co-ordinate space (Eqn. (\ref{form:lsm}) above), only atomic S- 
and P-states can
by mixed by it. The mixing matrix element has the form \cite{APVME}
\begin{eqnarray}
\label{obs:mix}
\bra{{\hbox{P}}}{\hat{\cal H}}^\sst{PV}_{\hbox{atom}}\ket{{\hbox{S}}} &=& 
(iG_F/2\sqrt{2}){\cal N}
C_{\hbox{sp}}(Z)\{ \qpv \\
&\ &\ \ +{\tilde k}{\cal Q}[F(F+1)-I(I+1)-3/4]\} \ \ \ , \nonumber
\end{eqnarray}
where ${\cal N}$ is a calculable normalization factor, $C_{\hbox{sp}}(Z)$ is an 
atomic
structure-dependent function, ${\cal Q}$ contains relativistic and nuclear 
finite size
corrections to the atomic wavefunctions at the origin, and $I$ and $F$ are the 
nuclear and atomic
angular momenta, respectively. The constant
${\tilde k}$ receives two contributions. One arises from the $V(e)\times A(f)$ 
$Z^0$-exchange
interaction. It is suppressed by the presence of $Q_\sst{W}^e=-1+4\sstw\sim 
-0.1$. The second
arises from the exchange of a photon between the electron an nucleus, where PV 
nucleon-nucleon
interactions generate an axial vector coupling of the photon to the nucleus. 
This coupling is
known as the nuclear anapole moment. It is proportional to the matrix element of 
the following
operator \cite{Hax89,Mus91}
\begin{equation}
\label{obs:ana}
{\tilde k}_{\hbox{anapole}}\propto\langle 0||\int\ d^3r\ r^2 
\left[J^\sst{EM}_\lambda+
\sqrt{2\pi}[Y_2\otimes J^\sst{EM}]_{1\lambda}\right]||0\rangle\ \ \ .
\end{equation}
Here, $\vec J^\sst{EM}$ is the nuclear EM current. Because of the factor of 
$r^2$ in the 
integrand of Eq. (\ref{obs:ana}), ${\tilde k}_{\hbox{anapole}}$ grows as 
$A^{2/3}$, where $A$ is
the nuclear mass number. For this reason, as well as the presence of 
$Q_\sst{W}^e$ in the
$V(e)\times A(f)$ $Z^0$-exchange atmplitude, the anapole moment contribution to 
the second term
of Eqn. (\ref{obs:mix}) is the dominant one for a heavy atom like cesium. The 
physics of the
anapole moment, and its connection to the PV hadronic weak interaction, is 
interesting in its own
right. It is discussed elsewhere in this book. 

As the presence of the angular momentum quantum numbers in Eq. (\ref{obs:mix}) 
implies, the first
term generates $\apvnsid$ and the second gives rise to $\apvnsd$. These two 
terms can be
separately determined by observing different atomic hyperfine transitions 
(different combinations
of
$F$ quantum numbers). For our purposes in this chapter, $\apvnsid$ is of primary 
interest. We may
write it as
\begin{equation}
\label{obs:nsid}
\apvnsid(N)=\xi\qpv = \xi\left[\qw0+Z\Delta\qwp+N\Delta\qwn\right]\ \ \ ,
\end{equation} 
where
\begin{equation}
\label{obs:qwatom}
\qw0 = Z(1-4\sstw)-N
\end{equation}
at tree level, where 
\begin{equation}
\label{obs:dqwatom}
Z\Delta\qwp+N\Delta\qwn = \zeta\left[(2Z+N)\hvu
	+(2N+Z)\hvd\right]\ \ \ ,
\end{equation}
where 
\begin{eqnarray}
\Delta\qwp&=&\zeta(2\hvu+\hvd)\\
\Delta\qwn&=&\zeta(2\hvd+\hvu)\ \ \ ,
\end{eqnarray}
as above and where $\xi$ is an atomic structure-dependent coefficient. A 
determination of
the latter generally requires theoretical knowledge of the relevant atomic
wavefunction and, therefore, introduces theoretical uncertainty into the
extraction of $\qpv$. The relative sensitivity of $\apv(N)$ to new physics
can be seen by rewriting $\qpv$ as
\begin{equation}
\label{obs:qwcor}
\qpv = \qw0\left[1+\delta_N\right]\ \ \ ,
\end{equation}
where
\begin{eqnarray}
\label{obs:delN}
\delta_N&=&(Z\Delta\qwp+N\Delta\qwn)/\qw0\\
&\approx& 
-\zeta\left[\left({2Z+N\over N}\right)\hvu
	+\left({2N+Z\over N}\right)\hvd\right]\nonumber\\
&=&-\zeta\left[(Z/N)(2\hvu+\hvd)
	+(2\hvd+\hvu)\right]\ \ \ \nonumber,
\end{eqnarray}
where the approximation $\qw0\approx -N$ has been made in light of the
small value for $1-4\sstw\approx 0.1$. From Eq. (\ref{obs:delN})
we observe that for
atoms having $Z\approx N$, the weak charge is roughly equally sensitive to
the new up- and down-quark vector current interactions.

The use of the isotope ratios involving $\apvnsid(N)$ and 
$\apvnsid(N')$ largely eliminates the dependence on the atomic 
structure-dependent
constant $\xi$ and the associated atomic theory uncertainty. I consider
two such ratios:
\begin{equation}
\label{obs:rat1}
{\cal R}_1 = {\apvnsid(N')-\apvnsid(N)\over\apvnsid(N')+\apvnsid(N)}
\end{equation}
and
\begin{equation}
\label{obs:rat2}
{\cal R}_2 = {\apvnsid(N')\over\apvnsid(N)}\ \ \ .
\end{equation}
To the extent that $\xi$ does not vary appreciably along the isotope chain,
one has
\begin{eqnarray}
\label{obs:rat1a}
{\cal R}_1 & = & {\qpv(N')-\qpv(N)\over \qpv(N')+\qpv(N)} \\
\label{obs:rat1b}
{\cal R}_2 & = & {\qpv(N')\over \qpv(N)} \ \ \ .
\end{eqnarray}
It is straightforward to work out the sensitivity of these ratios to
new physics. To this end, I write
\begin{equation}
\label{obs:ratcor}
{\cal R}_i = {\cal R}_i^0(1+\delta_i)\ \ \ ,
\end{equation}
where
\begin{eqnarray}
\label{obs:rat1sm}
{\cal R}_1^0 & = & {\qw0(N')-\qw0(N)\over \qw0(N')+\qw0(N)} \\
\label{obs:rat2sm}
{\cal R}_2^0 & = & {\qw0(N')\over \qw0(N)} \ \ \ ,
\end{eqnarray}
give the ratios in the Standard Model and the $\delta_i$ give corrections
arising from new physics. Letting $N'= N+\Delta N$ and dropping small
contributions containing $1-4\sstw$ one has
\begin{eqnarray}
\label{obs:rat1smapp}
{\cal R}_1^0 & \approx & {\Delta N\over 2N} \\
\label{obs:rat2smapp}
{\cal R}_2^0 & \approx & 1+{\Delta N\over N} 
\end{eqnarray}
and
\begin{eqnarray}
\label{obs:del1}
\delta_1 &\approx &  \zeta\left({2Z\over N+N'}
	\right)(2\hvu+\hvd)\\
\label{obs:del1}
\delta_2 &\approx & \zeta \left({Z\over N}\right)
	\left({\Delta N\over N'}\right)(2\hvu+\hvd) \ \ \ .
\end{eqnarray}

At first glance, the dependence of the $\delta_i$ $i=1,2$ on $\Delta\qwp=
\zeta(2\hvu+\hvd)$ and not $\Delta\qwn=\zeta(2\hvd+\hvu)$ may seem puzzling. 
As I point out in Ref. \cite{MRM98}, 
the shifts containing $\Delta\qwn$ in the numerator and
denominator of each ${\cal R}_i$ cancel to 
first order in $\zeta$. To illustrate, consider
${\cal R}_1$, for example, where one has
\begin{eqnarray}
\label{obs:ratnum}
\qpv(N')-\qpv(N)&\approx& -N'+N + (N'-N)\Delta\qwn\\
&=&(N-N')\left[1-\Delta\qwn\right] \nonumber
\end{eqnarray}
and
\begin{eqnarray}
\label{obs:ratden}
\qpv(N')+\qpv(N)&\approx&-(N+N') + (N+N')\Delta\qwn+2Z\Delta\qwp\\
&=&-(N+N')\left[1-\Delta\qwn-\left({2Z\over 
N+N'}\right)\Delta\qwp\right]\nonumber
\end{eqnarray}
so that in the ratio, the dependence on $\Delta\qwn$ cancels to first order. 
Because of this
feature, the ${\cal R}_i$ are twice as sensitive to new physics involving 
$u$-quarks than to
new physics which couples to $d$-quarks. The weak charge of a single isotope, on 
the
other hand, has essentially the same sensitivity to $u$- and $d$-quark new 
physics.

From a comparison of $\delta_N$ with the $\delta_i$, it is apparent that, for
a given experimental precision, the 
isotope ratios are generally less sensitive to direct new physics than is the 
weak
charge for a
single isotope. This statement is easiest to see in the case of
${\cal R}_2$, since $\delta_2$ contains the explicit factor $\Delta N/N'$.
Taking $Z\approx N$ for the case of ${\cal R}_1$, one finds that a single
isotope is three times more sensitive to new physics which couples to 
$d$-quarks and 1.5 times more sensitive to the $u$-quark coupling. For new
physics scenarios which favor new $e-d$ interactions over $e-u$ interactions,the 
weak charge for
a single isotope consititutes a more sensitive probe.

An alternative method for obtaining $\qpv$ is to scatter longitudinally
polarized electrons from fixed targets. Flipping the incident electron
helicity and comparing the helicity difference cross section with the
total cross section filters out the PV part of the weak neutral current
interaction. The resulting left-right asymmetry for elastic scattering
has the general form \cite{Mus94}
\begin{eqnarray}
\label{obs:alr}
\alr&=&{N_{+} - N_{-}\over N_{+} + N_{-}} \approx {2 M_\sst{NC}^\sst{PV}
\over M_\sst{EM}} \\
	&=& {G_F |q^2|\over 4\sqrt{2}\pi\alpha}\left[{\qpv\over Q_{EM}}
	+F(q)\right]\ \ \ .\nonumber
\end{eqnarray}
Here, $N_{+}$ ($N_{-}$) are the number of detected electrons for a positive
(negative) helicity incident beam; $M_\sst{EM}$ and $M_\sst{NC}^\sst{PV}$
are, respectively, the electromagnetic and parity-violating neutral current
electron-nucleus scattering amplitudes; $Q_{EM}$ is the nuclear EM charge;
and $F(q)$ is a correction involving hadronic and nuclear form factors. In
general, the latter term can be separated from the term containing the charges
by varying electron energy and angle. For elastic scattering, the charge
term can be isolated by going to forward angles and low energies. In the
case of PV M\" oller scattering, one has $F(q)\equiv 0$. The present PV
electron scattering program at MIT-Bates, Mainz-MAMI, and the Jefferson 
Laboratory seeks to determine the $F(q)$ for a variety of targets, with a
special emphasis on contributions from strange quarks \cite{McK89,Mue97,happex}.
The status and progress of this program is discussed in other chapters of this 
book.

In order to see how Eq. (\ref{obs:alr}) comes about, it is instructive to 
consider elastic
scattering from a positive parity, spin-zero, isospin-zero nucleus like $^4$He 
or $^{12}$C. For
this case, only the charge operator contributes to the scattering amplitudes. 
The weak charge
operator  for the nucleus is
\begin{equation}
\label{obs:qwhat}
{\hat Q}_\sst{W}=Q_\sst{W}^u u^{\dag} u +Q_\sst{W}^d d^{\dag} d +Q_\sst{W}^s 
s^{\dag} s + \cdots
\end{equation}
where $q^{\dag} q$ counts the number of quarks of flavor $q$ in the nucleus at 
zero momentum
transfer and where the $+\cdots$ represent the contributions from heavy quarks.
The weak charge operator can be re-expressed as follows:
\begin{eqnarray}
\label{obs:qwhatdecomp}
{\hat Q}_\sst{W}&=&(Q_\sst{W}^u-Q_\sst{W}^d) \frac{1}{2}(u^{\dag}u-d^{\dag}d)\\
&\ &\ 
+3(Q_\sst{W}^u+Q_\sst{W}^d)\frac{1}{6}(u^{\dag}u+d^{\dag}d-2s^{\dag}s)\nonumber\
\\
&\ &\ +(Q_\sst{W}^u+Q_\sst{W}^d+Q_\sst{W}^s)\ s^{\dag} s \ \ \ ,\nonumber
\end{eqnarray}
neglecting the three heaviest quark flavors.
The combination of currents in the first line is the isovector EM charge 
operator. Since the
nucleus has isospin zero, this combination cannot contribute to the scattering 
amplitude. The
combination in the second like is the isoscalar EM charge operator. Its 
contribution goes as
\begin{equation}
\label{obs:isoscal}
3(Q_\sst{W}^u+Q_\sst{W}^d) F_c(q)\ \ \ ,
\end{equation}
where $F_c(q)$ is just the EM form factor for the nucleus, with $F_c(q)=\qem=$ 
the nuclear
EM charge. The third term in Eq. (\ref{obs:qwhatdecomp}) contributes
\begin{equation}
\label{obs:strange}
(Q_\sst{W}^u+Q_\sst{W}^d+Q_\sst{W}^s) F_s(q)\ \ \ ,
\end{equation}
where $F_s(q)$ is the strange quark's vector current form factor for the 
nucleus. Since
stable nuclei have no net strangeness, one has $F_s(0)=0$ and 
$F_s(q)\sim q^2$ for small momentum transfer. The numerator in Eq. 
(\ref{obs:alr}) for $\alr$ is
just proportional to the sum of (\ref{obs:isoscal}) and (\ref{obs:strange}). The 
denominator is
propotional to
$F_c(q)$, with the same constant of proportionality. Consequently, in the ratio, 
$F_c(q)$ cancels
out of the first term, leaving $\alr$ proportional to the combination
\begin{equation}
\label{obs:qtot}
3(Q_\sst{W}^u+Q_\sst{W}^d) + (Q_\sst{W}^u+Q_\sst{W}^d+Q_\sst{W}^s) {F_s(q)\over 
F_c(q)}\ \ \ .
\end{equation}
This combination has just the form of the RHS of  Eqn. (\ref{obs:alr}), where 
the first term is
$q$-independent ratio of nuclear weak and EM charges and the second term is the 
$q$-dependent
form factor term. A similar line of reason applies to other targets. The only 
difference is that
when $J\not= 0$ and/or
$T\not=0$, the nucleus can support more form factors and the $F(q)$ term in the 
asymmetry
has a more complicated structure. 

In order to compare the sensitivities of different scattering experiments
to new physics, I specify the terms in Eq. (\ref{obs:alr}) for the following 
processes:
elastic scattering from the proton, $\alr(^1H)$; elastic scattering from
$(J^\pi, T)=(0^+,0)$, $\alr(0^+,0)$ nuclei; excitation of the $\Delta(1232)$
resonance, $\alr(N\to\Delta)$; and M\" oller scattering, $\alr(e)$.
The corresponding charge terms are (neglecting Standard Model radiative
corrections)
\begin{eqnarray}
\label{obs:ratprot}
\qpv(^1H)/Q_{EM}(^1H) & = & (1-4\sstw)\left[1+\delp\right] \\
\label{obs:ratzplus}
\qpv(0^+,0)/Q_{EM}(0^+,0) &=& -4\sstw\left[1+\delzp\right] \\
\label{obs:rate}
\qpv(e)/Q_{EM}(e) &=&(-1+4\sstw)\left[1+\dele\right]\ \ \ ,
\end{eqnarray}
while for the transition to the $\Delta$ one replaces the ratio of
charges by the ratio of isovector weak neutral current and EM couplings:
\begin{eqnarray}
\label{obs:ntod}
\qpv(N\to\Delta)/Q_{EM}(N\to\Delta)&\longrightarrow &2(1-2\sstw)
\left[1+\deld\right]\ \ \ .
\end{eqnarray}
The new physics corrections $\delta$ are given by
\begin{eqnarray}
\label{obs:delp}
\delp &=&\zeta(2\hvu+\hvd)/(1-4\sstw)\\
\label{obs:delz}
\delzp &=& -3\zeta(\hvu+\hvd)/(4\sstw)\\
\label{obs:dele}
\dele &=&-\zeta\hve/(1-4\sstw)\\
\label{obs:deld}
\deld &=&\zeta(\hvu-\hvd)/[2(1-2\sstw)]\ \ \ .
\end{eqnarray}

The expressions for the various $\delta_i$ allow us to make a few observations
regarding the relative sensitivities the corresponding observables to new 
physics. 
For this purpose, it is useful to take $\hvu=\hvd=1$ and specify $\delta_N$ for 
the case of
$^{133}$Cs. I also use cesium for the isotope rations and take a reasonable
range of neutron numbers:$N=78$, $N'=95$ \cite{Mas95}. In Table 1 I show the
$\delta_i$ in units of $\zeta$. The third column gives a scale factor $f$
defined as 
\begin{equation}
\label{obs:scalf}
f_i=\sqrt{\delta_i/\delta_N}\ \ \ .
\end{equation}
The factor $f_i$ can be used to scale the cesium APV limits on the new physics
mass scale $\Lambda$ to those obtainable from any other observable when measured
with the same precision as $\qpv({\hbox{Cs}})$: 
$\Lambda(i)=f_i\Lambda({\hbox{Cs}})$.
Alternatively, the limits from any other observable will be the same as those
from cesium when the precision is $f_i^2$ times the cesium uncertainty. 
The numbers shown in the Table are obtained using the $\bar{\hbox{MS}}$ value
$\sstw=0.232$ \cite{PDG98} while neglecting radiative corrections\footnote{The 
most
significant effect of radiative corrections appear in the M\" oller asymmetry
\cite{Cza96}.}.

\begin{table}[t]
\caption{\quad Relative sensitivities of PV observables to new physics, assuming $\hvu=\hvd$, tree-level values for the corresponding weak charges, and $\sin^2\hat{\theta}_\sst{W}=0.232$. The scale factor $f_i=\sqrt{\delta_i/\delta_N}$ can be used to scale mass bounds from the
cesium APV bounds to the bounds for observable $i$ assuming the same precision for 
both $\delta_N$ and $\delta_i$}

\begin{center}
\begin{tabular}{|c|c|}\hline
 & \\
 \hbox{Correction}$\quad \delta$& \hbox{Scale factor}$\quad f_i$\\  
  & \\ \hline
   & \\
 $\delta_N\approx 5.1\xi$ & 1 \\
  & \\
 $\delta_1\approx 1.9\xi$ & 0.6 \\
  &\\
 $\delta_2\approx 0.4\xi$ & 0.3 \\
  & \\
 $\delp\approx 40\xi$ & 2.8 \\
  & \\
 $\delzp\approx 6.5\xi$ & 1.1 \\
  & \\
 $\dele\approx 13\xi$ & 1.6  \\
  & \\
 $\delp = 0$ & 0 \\
  & \\ \hline
  \end{tabular}
\end{center}
\end{table}

As Table 1 illustrates, $\alr(ep)$ has by far the greatest generic 
sensitivity to
new physics for a given level of error in the observables. The reason is the
suppression of $Q_W^0$ for the proton, which goes as $1-4\sstw$ at tree level.
This suppression, however, renders the attainment of high precision more 
difficult
than for some of the other cases, since the statistical uncertainty in $\alr$
goes as $1/\alr$ \cite{Mus94,Mus92}. To set the scale, consider a 10\% 
$\alr(ep)$ measurement.
Using the scale factor $f_P$ in Table 1, we see that such a measurement 
would be roughly
comparable to the present cesium result in terms of new physics sensitivity. 
Given
the performance of the beam and detectors at the Jefferson Lab, it appears that 
a
future measurement of $\alr(ep)$ with  5\% or better precision could be feasible
\cite{Sou98}. Such a determination would yield new physics limits comparable to 
those
from cesium APV should the atomic theory error  be reduced to the level of the 
present
experimental error. A 2.5\% $ep$ measurement would strengthen the present APV 
bounds
by a factor of two. Constraints of this level would be competitive with those
expected from high energy colliders well into the next decade.
Similarly, a 0.8\% determination of the isotope ratio ${\cal R}_1$ would give 
new
physics limits comparable to the present cesium results. The prospects for 
achieving
this precision or better are promising. The Berkeley group, for example, expects 
to perform a
0.1\% determination of ${\cal R}_1$ using the isotopes of Yb $N=100\to N=106$ 
\cite{Bud98}. 
Similarly, the Seattle group plans to conduct studies on the isotopes of Ba$^+$ 
ions \cite{For98}.
For both Yb and Ba, the scale factors $f_1$ are similar to those for Cs, whereas 
$f_2$ depends
strongly on the range $\Delta N$. Note that no entry is listed for $f_\Delta$. 
Specifying
$\hvu=\hvd$ causes $\delta_\Delta$ to vanish. The $N\to\Delta$ asymmetry is 
sensitive only to new
physics having different  u- and d-quark interactions. As I illustrate in the 
following
section, deviations from this general pattern of relative sensitivities occur 
when specific new
physics scenarios are considered.

\section{Model Illustrations}
\label{sec:model}

With these general features in mind, we may now consider the implications of 
specific
models for new direct physics interactions. For illustrative purposes, I discuss 
three such scenarios: (a) the presence of a second \lq\lq low-mass" neutral 
gauge boson
in addition to the $Z^0$; (b) the possible existence of lepto-quarks; and (c) 
new interactions
arising from fermion sub-structure.

\medskip

\noindent {\bf A. Additional neutral gauge bosons.} 

The existence of one or more additional neutral gauge bosons is a natural 
consequence of
theories inspired by superstring theory. Certain versions of superstring theory 
depend on
the group structure E$_8\times$E$_8$ (for details, see Refs. 
\cite{Lon86,Moh92}). When
superstrings \lq\lq compactify" from 26 to four dimensions, one of these E$_8$ 
groups is reduced
to E$_6$. In the same way that the SU(2$)_L\times$U(1$)_Y$ gauge symmetry 
spontaneously breaks
down to U(1$)_\sst{EM}$ symmetry at the weak scale, giving the $W^\pm$ and $Z^0$ 
masses, it is
possible that the E$_6$ symmetry spontaneously breaks down to sub-group 
SO(N)$\times$ U(1) or 
SU(N)$\times$U(1) symmetries which ultimately break down to the 
SU(2$)_L\times$U(1$)_Y$ symmetry of the SM. At each occurence of spontaneous 
symmetry breakdown,
the gauge boson associated with the extra U(1) symmetry acquires a mass. It is 
possible
that at least one of the new gauge bosons is light enough to be observable at 
low-energies.
The models which generate such bosons from E$_6$ spontaneous symmetry breaking 
are known
as E$_6$ theories. 

An alternative reason to postulate the existence of additional $Z$-bosons is 
provided by
left-right symmetric theories. Such theories give a natural explanation for the 
observation
of PV  below the weak scale. According to the idea of LR symmetry, there exists 
an
SU(2$)_R$ gauge symmetry in addition to the SM electroweak gauge symmetry. The 
$W^\pm_R$ and
$Z_R$ bosons associated with this right-handed gauge group are much heavier than 
the SM
gauge bosons, so right handed-interactions are much weaker. However, the 
$W^\pm_R$ and
$Z_R$ may still be sufficiently light to generate very small corrections to 
low-energy
observables -- corrections which may become apparent when the experimental 
precision becomes
sufficiently high. 

If $Z'$ and $Z$ denote the \lq\lq new" and Standard Model neutral gauge bosons,
respectively, the existence of a light $Z'$  which mixes with the $Z$
is ruled out by $Z$-pole observables. The reason is that any new physics which 
mixes
with the $Z$ would show up strongly at the $Z$-pole. 
In the event that the $Z-Z'$ mixing
angle is $\approx 0$, however, LEP and SLC measurements provide rather
weak constraints. Consequently, I consider the case of zero mixing. 
For the sake of illustration, consider first the E$_6$ analysis of Ref. 
\cite{Lon86},
in which the different symmetry breaking scenarious can be parameterized
by writing the $Z'$ as
\begin{equation}
Z'= \cos\phi Z_\psi + \sin\phi Z_\chi \ \ \ .
\end{equation}
The $Z_\psi$ and $Z_\chi$ arise, for example, from the breakdown
E$_6\to$ SO(10)$\times$ U(1)$_\psi$ and SO(10)$\to$ SU(5)$\times $ U(1)$_\chi$.
Since the multiplets of SO(10) contain both $f$ and $\bar{f}$ for the
leptons and quarks of the Standard Model, $C$-invariance implies that the
$Z_\psi$ can have only axial vector couplings to these fermions. As a result,
it cannot contribute at tree-level to low-energy PV observables. In the case
of SU(5), however, the left-handed $d$-quark and $e^+$ live in a 
different multiplet from the left-handed $\bar{d}$ and $e^-$, whereas 
the $u$ and $\bar{u}$ live in the same multiplet. The $Z_\chi$ correspondingly
has both vector and axial vector couplings to the electron and $d$-quarks,
and only axial vector $u$-quark couplings. In short, E$_6$ $Z'$ bosons
yield $\hvu=0$ and $\hvd$, $\hve$ $\propto\sin\phi$.

According to the notation of Eq. (\ref{form:lnew}), we have for E$_6$ models
\begin{eqnarray}
\kappa^2&=&\alpha' \\
\Lambda^2&=& M_{Z'}^2\\
\hvu&=&0 \\
\label{model:E6}
\hvd=-\hve&=&\left[\sqrt{15}\sin\phi\cos\phi/3-\sin^2\phi\right]\ \ \ ,
\end{eqnarray}
where $\alpha'$ is the fine structure constant associated with the new
gauge coupling. Generally, one has 
\begin{equation}
\label{model:ap}
\alpha'\simle{5\over 3} {\alpha\over\cos^2\theta_\sst{W}}\approx 2.2\alpha
\ \ \ .
\end{equation}

Different models for the $Z'$ correspond to different choices for
$\phi$. Examples include the $Z_\eta$ ($\tan\phi=-\sqrt{3/5}$) and
the $Z_I$ ($\tan\phi=-\sqrt{5/3}$), where the latter is associated
with an additional \lq\lq inert" SU(2) gauge group not contributing to 
the electromagnetic charge. From the
standpoint of phenomenology, it is  worth noting the dependence of $\hvd$ and 
$\hve$ on the value
of $\phi$. For $\phi=\phi_c=\tan^{-1}(\sqrt{5/3})$, $\hvd=0=\hve$. For 
$\phi>\phi_c$,
$\hvd>0$. In the event that any of the $\delta_i$ is determined to be non-zero, 
the sign
of the deviation would constrain the allowable range of E$_6$ models. 
 From Eq. (\ref{obs:delN}), we observe, for example,  that $\delta_N$ is 
negative for
$\hvu=0$ and $\hvd>0$. At the $1\sigma$ level, the most recent value of  
$\delta_N$ for cesium
implies that $\hvd>0$, and therefore could not be explained models
giving $\phi<\phi_c$.  The model which gives the largest possible contribution 
to
the weak charge is the $Z_\chi$, which corresponds to $\phi=90^\circ$.

In left-right symmetric theories, the low energy
gauge group becomes SU(2$)_L\times$SU(2$)_R\times$U(1$)_{B-L}$, where 
$B-L=1/3$ for baryons and $-1$ for leptons. In the case of \lq\lq manifest"
left-right symmetry the SU(2$)_L$ and SU(2$)_R$ couplings are identical.
For this case, a second low-mass neutral gauge boson $Z_{R}$ couples to
fermions with the strength \cite{Lan92,Moh92}
\begin{eqnarray}
\hvu&=&-\frac{3}{5}\frac{\alpha}{4}\left(\frac{\alpha}{4}-\frac{1}{6\alpha}
\right)\\
\hvd&=&\frac{3}{5}\frac{\alpha}{4}\left(\frac{\alpha}{4}+\frac{1}{6\alpha}\right
)
\\
\hvu&=&\frac{3}{5}\frac{\alpha}{4}\left(\frac{\alpha}{4}-\frac{1}{2\alpha}\right
)
\end{eqnarray}
where 
\begin{equation}
\alpha=\left({1-2\sstw\over\sstw}\right)^{1/2}\approx 1.53\ \ \ .
\end{equation} 
With this set of couplings, the combination appearing in the correction to the
proton's weak charge is $2\hvu+\hvd\approx 0.012 << \hvu, \ \hvd$. Consequently,
the sensitivies of the $R_i$ and $\alr(^1{\hbox{H}})$ are suppressed relative
to their generic scale. The corresponding mass limits on $M_{Z_{R}}$ are weaker
than those obtainable from cesium APV or $\alr(0^+,0)$. 

In Table 2, I give the present and prospective for two species of additional
neutral gauge bosons, the $Z_\chi$ and $Z_{R}$. In particular, 
we show lower bounds on the Fermi constant associated
with the new gauge boson $Z'$, defined as
\begin{equation}
\label{model:gfnew}
{G_a^\prime\over \sqrt{2}} = {g^{\prime\ 2}\over 8 M_{Z^\prime_a}^2}\ \ \ ,
\end{equation}
where $g^\prime$ is the coupling associated with the additional $U(1)_a$ gauge 
group.
Low-energy PV observables constrain the ratio $g^\prime/M_{Z^\prime_a}$ and do 
not provide
separate limits on the mass and coupling. Consequently, the ratio of 
$G^\prime_\chi/G_F$
characterizes the strength of a new $U(1)_\chi$ gauge interaction relative to 
the strength
of the Standard Model. In general, mass bounds for the $Z'$ can be obtained from 
the limits on
$G'$ under specific assumptions for $g'$. A comparison of such mass bounds is 
often
instructive, so I quote such bounds in the final two columns of Table 2.  
Lower bounds on $M_\chi$ are quoted assuming the
maximal value for $g'$ as given by Eq. (\ref{model:ap}). In the case of LR 
symmetry models with
manifest LR symmetry, one has $g'=g$. The corresponding mass limits for the 
$Z_{R}$ are given
in the final column of Table 2. Since I only discuss the case of manifest LR 
symmetry
above, I do not include bounds on $G^\prime_{R}/G_F$.
\begin{table}[t]
\caption{\quad Present and prospective limits on
two species of additional neutral gauge bosons. The third column gives the ratio
of fermi constants as defined in the text. The fourth and fifth columns give 
lower bounds on masses for the $Z_\chi$ and $Z_{R}$, respectively, assuming the
precision given in column two. The M\"oller limits are derived without 
accounting
for Standard Model radiative corrections.}
\vspace{0.2cm}
\begin{center}
\footnotesize
\begin{tabular}{|c|c|c|c|c|}\hline 
 & & & & \\
 \hbox{Observable} & \hbox{Precision} & $G^\prime_\chi/G_f$ & $M_{Z_\chi}\ 
({\hbox{GeV}})$&$M_{Z_{R}}\ ({\hbox{GeV}})$\\ 
& & & & \\ \hline
& & & & \\
$\qpv({\hbox{Cs}})$ & 1.3\% & 0.006 & 730 & 790 \\
& & & & \\
  & 0.35\%  & 0.0016 & 1410 & 1520 \\
 & & & & \\
 ${\cal R}_1$ & 0.3\% & 0.006 & 740 & 360 \\
 & & & & \\
  & 0.1\% & 0.002 & 1300 & 630 \\
 & & & & \\ 
$\qpv(^1{\hbox{H}})/Q_\sst{EM}(^1{\hbox{H}})$ & 10\% & 0.010 & 580 & 285 \\ 
 & & & &\\
 & 3\% & 0.003 & 1100 & 520 \\
  & & & &\\
 $\qpv(0^+,0)/Q_\sst{EM}(0^+,0)$& 1\% & 0.004 & 910 & 920 \\
  & & & &\\
 $\qpv(e)/Q_\sst{EM}(e) $& 7\% & 0.013 & 700 & 350  \\
 & & & & \\
 $\alr(N\to\Delta)$& 1\% & 0.013 & 490 & 920 \\
 & & & & \\ \hline
\end{tabular}
\end{center}
\end{table}

The limits in Table 2 lead to several observations. Primary among these is 
that
low-energy PV already constrains the strength of new, low-energy gauge 
interactions
to be at most a few parts in a thousand relative to the strength of the 
SU(2$)_L\times$U(1$)_Y$ sector. When reasonable assumptions are made about new 
gauge
couplings strengths, low-energy mass bounds now approach one TeV. The 
significance of
these bounds becomes more apparent when a comparison is made with the results of 
collider
experiments. The present 110 pb$^{-1}$ $p\bar p$ data set analyzed by the CDF 
collaboration
yields a lower bound on $M_{Z_{R}}$ of 620 GeV, assuming manifest LR symmetry 
\cite{Abe97a}. The
lower bound for $M_{Z_\chi}$ is 585 GeV, assuming no $Z_\chi$ decays to 
supersymmetric particles.
The sensitivity of cesium APV already exceeds that of the Tevatron experiments. 
I wish to
emphasize that collider experiments and low-energy PV provide complementary 
probes of extended
gauge group structure. PV observables are sensitive to the vector couplings of 
the $Z'$ to
fermions. For a model for which this coupling is small or vanishing ({\em e.g.}, 
the $Z_\psi$
having
$\phi=0^\circ$ in Eq. (\ref{model:E6})), PV observables will yield no 
significant information.
Collider experiments, on the other hand, retain a sensitivity to such $Z'$ 
interactions. For
models in which the
$ffZ'$ coupling is not suppressed, low-energy PV yields the most stringent 
bounds.

A look to the future suggests that PV could continue to play such a 
complementary role. 
Assuming the collection of 10 fb$^{-1}$ of data at TeV33, for
example, the current Tevatron bounds on $M_{Z'}$ would increase by roughly a 
factor of
two \cite{Riz96b}. The prospective sensitivity of cesium APV, assuming a 
reduction in atomic
theory error to the level of the present experimental uncertainty, would exceed 
the collider
sensitivity by
$\sim$ 50\%. Precise determinations of the isotope ratio ${\cal R}_1$ or various 
PV electron
scattering asymmetries could also yield sensitivities which match or exceed the 
prospective
TeV33 reach. Only with the advent of the LHC or $\simge 60$ TeV hadron collider 
will high-energy
machines probe masses significantly beyond those accessible with low-energy PV. 

Finally, Table 2 illustrates the model-sensitivity of different PV 
observables. For the
models considered here, the mass bounds do not scale with the $f_i$ of Table 1 
since 
$\hvu\not=\hvd$. Both the $Z'$ in $E_6$ and the $Z_{LR}$ couple more strongly to 
protons
than neturons. Consequently, both $R_1$ and $\alr(^1{\hbox{H}})$ display weaker 
sensitivity
to new gauge interactions than their generic sensitivies to new physics 
indicated in Table 1.

\bigskip
\noindent{\bf B. Leptoquarks}

\medskip
In early 1997, the H1 \cite{Adl97} and ZEUS \cite{Bre97} collaborations reported 
the presence
of anomalous events in high-$|q^2|$ $e^+p$ collisions at HERA. These
events have been widely interpreted as arising from s-channel lepton-quark
resonances with mass $\mlq \approx 200$ GeV. Such so-called leptoquarks are
particles, either scalar or vector, which carry both lepton and baryon number.
In $e^+p$ collisions, a leptoquark would be formed in the $s$-channel when 
the $e^+$ and quark annihilate into a leptoquark, which subsequently decays
back into the $e^+q$ pair (see Fig 2). The stringent limits
on the existence of vector leptoquarks (LQ's) obtained at Fermilab make scalar
leptoquarks  the favored interpretation of the HERA events. Although
the results remain ambiguous and are open to alternative explanations,
they are nonetheless provacative and suggest a consideration of LQ effects
in low-energy PV processes. To that end, I consider general LQ interactions
of the form
\begin{eqnarray}
\label{model:LQscal}
{\cal L}_\sst{LQ}^\sst{S}&=& \lscal(\phi{\bar e}_L q_R+{\hbox{h.c.}})\\
\label{model:LQvect}
{\cal L}_\sst{LQ}^\sst{V}&=& \lvect({\bar e}_L\gamma_\mu q_L\phi^\mu
	+{\hbox{h.c.}})
\end{eqnarray}
where $\phi$ and $\phi^\mu$ denote scalar and vector LQ fields, respectively.
For simplicity, we do not explicitly consider the corresponding interactions
obtained from Eqs. (\ref{model:LQscal}, \ref{model:LQvect}) with 
$L\leftrightarrow R$. The
corresponding analysis is similar to what follows. Assuming $\mlqs >> |q^2|$, 
the processes of
Fig. 2 give rise to the following PV interactions:
\begin{eqnarray}
\label{model:lscala}
{\cal L}^\sst{S}_\sst{PV}&=&(\lscal/2\mlq)^2 \left[{\bar e}q {\bar q}\gamma_5 e
	-{\bar e}\gamma_5 q{\bar q} e\right]\\
\label{model:lvecta}
{\cal L}^\sst{V}_\sst{PV}&=&(\lvect/2\mlq)^2\left[{\bar e}\gamma_\mu q{\bar q}
\gamma^\mu \gamma_5 e+{\bar e}\gamma_\mu \gamma_5 q{\bar q}\gamma^\mu e\right]
\end{eqnarray}
After a Fierz transformation, these become
\begin{eqnarray}
\label{model:lscalb}
{\cal L}^\sst{S}_\sst{PV}&=&(\lscal/2\sqrt{2}\mlq)^2[{\bar e}\gamma_\mu
	\gamma_5 e{\bar q}\gamma^\mu q -{\bar e}\gamma_\mu e {\bar q}
	\gamma^\mu\gamma_5 q\\
\label{model:lvectb}
	&& +{1\over 4} {\bar e}\sigma_{\mu\nu} e {\bar q}\sigma^{\mu\nu}
	\gamma_5 q - {1\over 4}{\bar e}\sigma_{\mu\nu}\gamma_5 e
	{\bar q}\sigma_{\mu\nu} q]\nonumber\\
&&\nonumber \\
{\cal L}^\sst{V}_\sst{PV}&=&-(\lvect/2\mlq)^2[{\bar e}\gamma_\mu e
	{\bar q}\gamma^\mu\gamma_5 q +{\bar e}\gamma_\mu\gamma_5 e
	{\bar q}\gamma^\mu q]
\end{eqnarray}
In terms of the interaction in Eq. (\ref{form:lnew}), we may identify
\begin{eqnarray}
\Lambda^2&=&\mlqs\\
\kappa^2&=&\lambda^2/16\pi
\end{eqnarray}
and $\hvq=1/2$ ($\hvu=-1$) for scalar (vector) LQ interactions.

Assuming for simplicity that either a u-type or d-type LQ (but not
both) contributes to low-energy PV processes, the results from cesium
APV, together with Eqs. (\ref{model:lscalb}) and (\ref{model:lvectb}), yield the 
following
$1\sigma$ limits on LQ couplings and masses:
\begin{equation}
\label{model:lscallim}
\lscal\leq\cases{0.042\  (\mlq/100\ \mbox{GeV})\ , & u-type \cr
&\cr
0.04\ (\mlq/100\ \mbox{GeV})\ , & d-type\cr}
\end{equation}
and
\begin{equation}
\lvect\leq\cases{0.030 (\mlq/100\ \mbox{GeV}), & u-type\cr
&\cr
0.028(\mlq/100\ \mbox{GeV}), & d-type\cr} \ \ \ .
\end{equation}

The HERA results are rather insensitive to the value of the coupling $\lscal$.
Substituting the HERA value of $\approx$ 200 GeV into (\ref{model:lscallim}) 
yields an
upper bound of $\lscal\simle 0.08$. On general grounds, one might have
expected $\kappa^2\sim\alpha$ or $\lscal\sim 0.6$. The cesium APV results
require the coupling for a 200 GeV scalar LQ to be about an order of
magnitude smaller than this expectation. Alternatively, if one does
not interpret the HERA results as a 200 GeV LQ and assumes $\kappa^2
\sim\alpha$, the APV bounds on the scalar LQ mass are $\mlq>$ 1.5 TeV. 
\begin{figure}
\begin{center}
\epsfbox{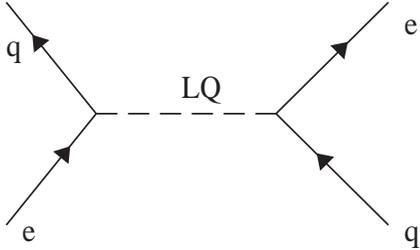}
\caption{\label{Fig2} Parity-violating semi-leptonic amplitude generated by 
leptoquark
(LQ) exchange.}
\end{center}
\end{figure}
Table 3 gives comparable bounds on the LQ coupling-to-mass ratio for
the other PV observables discussed in Section II. The bounds are characterized
by the quantity $\gamma_q$, defined as
\begin{equation}
\label{model:gamq}
\lscal \leq \gamma_q\ (\mlq/100\ \mbox{GeV}) \ \ \ ,
\end{equation} 
where $q$ denotes the quark flavor.

\begin{table}[t]
\caption{\quad Present and prospective limits on
leptoquark interactions. Third and fourth columns give $\gamma_q$ for a
$q$-type leptoquark, as defined in Eq. (\ref{model:gamq}). Leptoquark 
sensitivity of 
M\"oller asymmetry does not behave according to Eq. (\ref{model:gamq}), so that 
no limits
on the $\gamma_q$ are attainable.}
\vspace{0.2cm}
\begin{center}
\footnotesize
\begin{tabular}{|c|c|c|c|} \hline
 & & & \\
$\hbox{Observable}$ & $\hbox{Precision}$ & $\gamma_u$ & $\gamma_d$ \\
 & & & \\ \hline
 & & & \\
 $\qpv({\hbox{Cs}})$ & 1.3\% & 0.04 & 0.042  \\
 & & & \\
  & 0.35\%  &  0.021 & 0.022 \\
  & & & \\
 ${\cal R}_1$ & 0.3\% & 0.04 & 0.028 \\
 & & & \\
 &  0.1\% & 0.023 & 0.016  \\
 & & & \\
 $\qpv(^1{\hbox{H}})/Q_\sst{EM}(^1{\hbox{H}})$ & 10\% & 0.05 & 0.036 \\
 & & & \\
 & 3\% & 0.028 & 0.02  \\
 & & & \\
 $\qpv(0^+,0)/Q_\sst{EM}(0^+,0)$ & 1\% & 0.033 & 0.033 \\
 & & & \\
 $\qpv(e)/Q_\sst{EM}(e)$ & 7\% & $--$ & $--$ \\
 & & & \\
 $\alr(N\to\Delta)$ & 1\% & 0.06 & 0.06 \\
& & & \\ \hline
\end{tabular}
\end{center}
\end{table}

Note that no bounds
are given for the M\" oller asymmetry, as LQ's do not contribute at
tree level. As shown in Ref. \cite{MRM98}, the leading contributions arise from 
the loop graphs of
Fig. 3.  The corresponding contributions to the PV effective $ee$ interaction 
are
\begin{eqnarray}
\label{model:loopa}
M^\sst{PV}_{(a)}&=& \left({\lambda_\sst{S}^2\over 16\pi\mlq}\right)^2{\bar 
e}\gamma_\mu e
	{\bar e}\gamma^\mu\gamma_5 e \\
\label{model:loopb}
M^\sst{PV}_{(b)}&=& {\alpha Q_q\over 12\pi}\left({\lscal\over\mlq}\right)^2
	\ln {m_q\over\mlq} {\bar e}\gamma_\mu e{\bar e}\gamma^\mu
	\gamma_5 e
\end{eqnarray}
where $m_q$ and $Q_q$ are the intermediate state quark mass and
E.M. charge. For $\mlq=100$ GeV, a 7\% determination of the M\" oller
asymmetry would yield
\begin{equation}
\lscal\leq 1.2
\end{equation}
from graph (a) and 
\begin{equation}
\lscal\leq
\cases{1.14, & d-type \cr
	      0.78, & u-type\cr}
%\begin{array}{c2}
%6.1 & \mbox{d-type}\\
%4.3 & \mbox{u-type}
%\end{array}\right
\end{equation}
from graph (b). The limits for a vector LQ are comparable. The prospective
M\" oller bounds are about ten times weaker than those attainable with
semi-leptonic PV. Any deviation of the M\" oller asymmetry from the
Standard Model prediction is unlikely to be due to LQ's. 

%\medskip
\begin{figure}
\begin{center}
\epsfbox{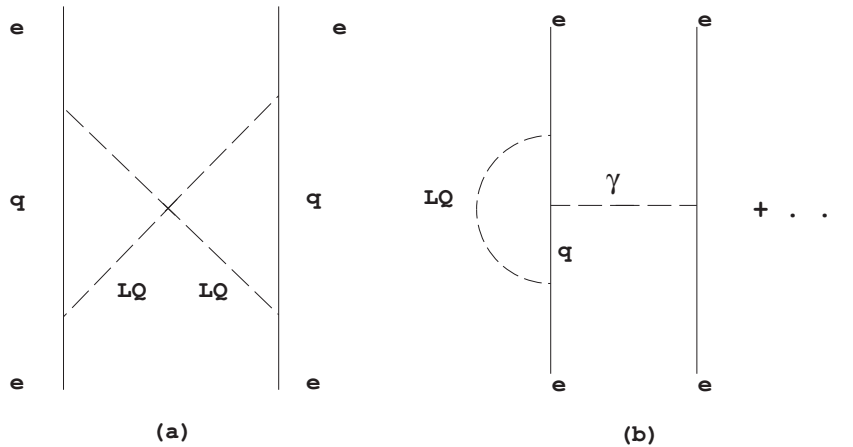}
\end{center}
\caption{\label{Fig 3} Leptoquark (LQ) one-loop contributions to PV M\"oller 
scattering.}
\end{figure}
\noindent{\bf C. Compositeness}

\medskip

The Standard Model assumes the known bosons and fermions to be pointlike.
The possibility that they possess internal structure, however, remains an
intriguing one. Manifestations of such composite structure could include
the presence of fermion form factors in elementary scattering processes
\cite{Abe97b} or the existence of new, low-energy contact interactions 
\cite{Eic83}. The
latter could arise, for example, from the interchange of fermion
constituents at very short distances~\cite{Lan92}.  A recent analysis of
$p{\bar p}\to \ell^+\ell^-$ data by the CDF collaboration  limits the
size of a lepton or quark to be $R< 5.6\times 10^{-4}$ f when $R$ is
determined from the assumed presence of a form factor at the
fermion-boson vertex\,\cite{Abe97b}. More stringent limits on the distance scale
associated with compositeness are obtained from the assumption of new
contact interations governed by a coupling of strength $g^2=4\pi$.
 Collider experiments yield $R\sim 1/\Lambda < 6
\times 10^{-5}$ f, where $\Lambda$ is the mass scale associated with new
dimension six lepton-quark operators~\cite{Abe97b}.

It is conventional to write the lowest dimenion contact interactions as
\begin{equation}
\label{model:lcompa}
{\cal L}_\sst{COMP} = 4\pi\sum_{ij}{\eta_{ij}\over\Lambda_{ij}^2} 
{\bar e}_i \gamma_\mu e_i {\bar q}_j\gamma^\mu q_j
\ \ \ ,
\end{equation}
where $\Gamma$ is any one of the Dirac matrices and $i,j$ denote the
appropriate fermion chiralities ({\em e.g.}, ${\bar e}_L\gamma_\mu e_L {\bar 
q}_R 
\gamma^\mu q_R$ {\em etc.}).
The quantities $\eta_{ij}$ take on the values $\pm 1, 0$ depending on
one's model assumptions. In terms of the PV interaction of Eq. 
(\ref{form:lnew}), the
contribution from ${\cal L}_\sst{COMP}$is
\begin{equation}
\label{model:lcompb}
\pi {\bar e}\gamma_\mu \gamma_5 e
\sum_{q}\left[{\eta_{RR}\over\Lambda_{RR}^2}-{\eta_{LL}\over\Lambda_{LL}^2}
+{\eta_{RL}\over\Lambda_{RL}^2}-{\eta_{LR}\over\Lambda_{LR}^2}\right]
{\bar q}\gamma^\mu q\ \ \ .
\end{equation}  
Writing this interaction in terms of a common mass scale $\Lambda$ yields
\begin{equation}
\label{model:lcompc}
{\pi\over\Lambda^2}{\bar e}\gamma_\mu\gamma_5 e\sum_q\left[
{\tilde\eta}_{RR}-{\tilde\eta}_{LL}+{\tilde\eta}_{RL}-{\tilde\eta}_{LR}
\right] \ \ \ , 
\end{equation}
where
\begin{equation}
{\tilde\eta}_{ij}=\eta_{ij}\left({\Lambda\over\Lambda_{ij}}\right)^2
\ \ \ .
\end{equation}
The correspondence with ${\cal L}_\sst{NEW}^{PV}$ is given by
\begin{eqnarray}
\kappa^2&=&1/4 \\
\hvq&=&{\tilde\eta}_{RR}-{\tilde\eta}_{LL}+{\tilde\eta}_{RL}-
{\tilde\eta}_{LR}
\end{eqnarray}
It is worth noting that in the two previous model illustrations, one had 
$\kappa^2\sim\alpha$
or smaller. The present case with $\kappa^2$ on the order of unity corresponds 
to a new strong
interaction. 

On the most general grounds, one has no strong argument for any of the
$\hvq$ to vanish. Consequently, low energy observables will generate lower
bounds on $\Lambda$. To compare with the recent $p\bar{p}$ collider limits, 
consider the
case of ${\tilde\eta}_{LL}=\pm 1$ and ${\tilde\eta}_{RR}={\tilde\eta}_{RL}=
{\tilde\eta}_{LR}=0$.In this case, the cesium APV results yield
\begin{equation}
\label{model:lcomplim}
\Lambda_{LL}\geq 17.3 \ \ {\hbox{TeV}}
\end{equation}
assuming $\hvu=\hvd=-{\tilde\eta}_{LL}$.  Regarding other low-energy PV
observables, we note that the general comparisons made in Section III apply
here. For example, a 10\% measurement of $\delta_P$ with PV $ep$ scattering
would yield comparable bounds, while a measurement of the isotopte ration
${\cal R}_1$ with 0.8\% precision would be required to obtain comparable limits.
Were the cesium APV theory error reduced to the level of the present 
experimental
error, or were a 2\% determination of $\delta_P$ achieved, the lower limit 
(\ref{model:lcomplim})
would double.  

Specific bounds from present and prospective measurements are given in Table 4 
below:
\begin{table}[t]
\caption{\quad Present and prospective limits on
compositeness scale for the \lq\lq LL" scenario. (a) M\"oller limits refer
to new $ee$ compositeness interactions, while other enteries refer to 
$eq$ interactions.}
\vspace{0.2cm}
\begin{center}
\footnotesize
%\begin{tabular}
$$\hbox{\vbox{\offinterlineskip
\def\strut{\hbox{\vrule height 15pt depth 10pt width 0pt}}
\hrule
\halign{
\strut\vrule#\tabskip 0.2cm&
\hfil$#$\hfil&
\vrule#&
%\hfil$#$\hfil&
%\vrule#&
%\hfil$#$\hfil&
%\vrule#&
\hfil$#$\hfil&
\vrule#&
\hfil$#$\hfil&
\vrule#\tabskip 0.0in\cr
%& \multispan5{\hfil\bf TABLE IV.3\hfil} & \cr\noalign{\hrule}
& \hbox{Observable} && \hbox{Precision} && \Lambda_{LL}\ ({\hbox{TeV}}) & 
\cr\noalign{\hrule}
& \qpv({\hbox{Cs}}) && 1.3\% && 17.3 & \cr
& && 0.35\% && 33.3 &\cr
& {\cal R}_1 &&  0.3\% && 21.6 &\cr
& && 0.1\% && 37.4 &\cr
& \qpv(^1{\hbox{H}})/Q_\sst{EM}(^1{\hbox{H}}) && 10\% && 17.5&\cr
& && 3\% && 31.8 &\cr
& \qpv(0^+,0)/Q_\sst{EM}(0^+,0) && 1\% && 21.6 &\cr
& \qpv(e)/Q_\sst{EM}(e) && 7\% && 13.1^{\hbox{a}} &\cr
\noalign{\hrule}}}}$$
%\end{tabular}
\end{center}
\end{table}

As with other new physics scenarios, the present and prospective low-energy 
limits
on compositeness are competitive with those presently obtainable from collider
experiments as well as those expected in the future. The CDF collaboration has
obtained lower bounds on $\Lambda_{LL}(eq)$ of 2.5 (3.7) TeV for  
${\tilde\eta}_{LL}=+1\ 
(-1)$ \cite{Abe97b}. One expects to improve these bounds to 6.5 (10) TeV with 
the completion
of Run II  and 14 (20) TeV with TeV33 \cite{Riz96}. It is conceivable that 
future improvements
in determinations of $\qpv$ with APV or scattering will yield stronger bounds 
that those
expected from colliders. In the case of $\Lambda_{LL}(ee)$, $Z$-pole observables 
imply
lower bounds of 2.4 (2.2) TeV for ${\tilde\eta}_{LL}=+1\ (-1)$ \cite{Ack97}. 
The prospective M\"oller PV lower bounds exceed the LEP limits considerably. 

\section{Theoretical Uncertainties}
\label{sec:theo}

The attainment of stringent limits on new physics scenarios from low-energy 
PV requires that conventional many-body physics of atoms and hadrons be 
sufficiently
well understood. At present, the dominant uncertainty in
$\qpv({\hbox{Cs}})$ is theoretical. A significant improvement in the precision 
with which this
quantity is known would rely on corresponding progress in atomic theory. The 
issues involved
in reducing the atomic theory uncertainty are discussed elsewhere in this book. 
In what follows,
I discuss the many-body uncertainties associated with the other observables 
discussed above.

\medskip
\noindent{\bf A. Isotope ratios}
\medskip

It was pointed out in Refs. \cite{For90,Pol92} that the isotope ratios ${\cal 
R}_i$ display an
enhanced sensitivity to the neutron distribution $\rho_n(r)$ within atomic 
nuclei, and that
uncertainties in $\rho_n(r)$ could hamper the extraction of new physics limits 
from the ${\cal
R}_i$.  To see how this $\rho_n$-sensitivity comes about, I follow Refs. 
\cite{For90,Pol92}
and consider a simple model in which the nucleus is treated as a sphere of 
uniform proton and
neutron number densities out to radii
$R_P$ and $R_N$, respectively. In this case, one may express the weak charge as
\begin{equation}
\label{theo:qw}
\qpv= Z \qwp q_p + N\qwn q_n \ \ \ ,
\end{equation}
where
\begin{eqnarray}
\label{theo:qp}
q_p&=& (1/{\cal N})\int\ d^3x \bra{P}{\hat\psi}_e^{\dag}({\vec 
x})\gamma_5{\hat\psi}_e({\vec x})
\ket{S}\rho_p({\vec x})\\ 
\label{theo:qn}
q_n&=& (1/{\cal N})\int\ d^3x \bra{P}{\hat\psi}_e^{\dag}({\vec
x})\gamma_5{\hat\psi}_e({\vec x})\ket{S}
\rho_n({\vec x})
\end{eqnarray}
where ${\hat\psi}_e({\vec x})$ is the electron field operator, $\ket{S}$ and 
$\ket{P}$ are
atomic $S_{1/2}$ and $P_{1/2}$ states, and  ${\cal N}$ is the value of the 
electron matrix
element at the origin. The latter matrix element may be written as
\begin{equation}
\label{theo:pscal}
\bra{P}{\hat\psi}_e^{\dag}({\vec x})\gamma_5{\hat\psi}({\vec x})\ket{S}={\cal N} 
f(x) \ \ \ ,
\end{equation}
where $f(0)=1$. The effect of uncertainties in $\rho_p({\vec x})$ -- which are 
smaller than those
in $\rho_n({\vec x})$ -- are suppressed in $\qpv$ since $q_p$ is multiplied by 
the small
number $\qwp$. Consequently, we consider only $q_n$. In the simple nuclear model 
discussed 
above, one obtains
\begin{equation}
\label{theo:qnexp}
q_n=1-(Z\alpha)^2 f_2^N+\cdots\ \ \ ,
\end{equation}
where
\begin{eqnarray}
f_2^N&=& {3\over 10}x_N^2 -{3\over 70} x_N^4 +{1\over 450} x_N^6\\
x_N&=& R_N/R_P \ \ \ .
\end{eqnarray}
Letting $\delta^n_N$ and $\delta^n_i$ denote the $\rho_n({\vec x})$ corrections 
to $\qpv(N)$ and
${\cal R}_i$, respectively, we obtain
\begin{eqnarray}
\delta^n_N&\approx& -(Z\alpha)^2 f_2^N(x_N)\\
\delta^n_1&\approx& -(Z\alpha)^2 (N'/\Delta N) f_2^{N\prime}(x_N)\Delta x_N\\
\delta^n_2&\approx& -(Z\alpha)^2 f_2^{N\prime}(x_N) \Delta x_N\ \ \ ,
\end{eqnarray}
where $\Delta x_N=(R_{N'}-R_N)/R_P$. Uncertainties in $\qpv$ and ${\cal R}_i$ 
arise from
{\em uncertainties} in the foregoin quantities. As shown in Ref. \cite{MRM98}, 
if measurements of the ${\cal R}_i$ are to generate new physics
limits comparable to those obtainable from $\qpv({\hbox{Cs}})$ ( assuming a 
future reduction
of atomic theory error to the experimental error level) one will need 
$\delta(\Delta x_N)\leq
0.004$ for ${\cal R}_1$ and $\delta (\Delta x_N)\leq 0.02$ for ${\cal R}_2$. 
These requirements
on $\Delta x_N$ will differ for other atoms, such as Yb or Ba, depending on the 
values of $N$
and $\Delta N$.  At present, there exist no reliable experimental determinations 
of $x_N$ or
$\Delta x_N$, so that the interpretation of APV observables must rely on nuclear 
theory. It is
conceivable that the theory uncertainty in $x_N$ is 5\% or better 
\cite{Pol92,Che93}.
Consequently, one could argue that even if the atomic theory error in 
$\qpv({\hbox{Cs}})$ were
reduced to the present experimental error, neutron distribution uncertainties 
should not
complicate the extraction of new physics constraints. The situation regarding 
isotope shifts is
more debatable.

Given the present theoretical situation, a model-independent determination of 
$\rho_n({\vec x})$
is desirable. To that end, PVES may prove useful \cite{Don89}. The basic idea is 
that the $F(q)$
term in Eq. (\ref{obs:alr}) depends on the fourier transform of $\rho_n(r)$ 
among other things. By
measuring the $q$-dependence of $\alr$, then, one may be able to extract enough 
information on
$\rho_n(r)$ to reduce the corresponding uncertainties in the ${\cal R}_i$. To 
illustrate the
possibility, I consider a
$(J^\pi, T)=(0^+,0)$ nucleus, such as $^{138}_{56}$Ba, noting that the isotopes 
of barium
are under consideration for future APV isotope ratio measurements. As noted in 
Refs.
\cite{Don89,Mus94}, the PV asymmetry for $(0^+,0)$ nuclei may be written as
\begin{equation}
\label{theo:rhona}
-\left[{4\sqrt{2}\pi\alpha\over G_F|q^2|}\right]\alr = \qwp+\qwn{\int\ d^3x\ 
j_0(qx) 
\rho_n({\vec x}) \over \int\ d^3x\ j_0(qx) \rho_p({\vec x})}\ \ \ .
\end{equation}
Since $|\qwp/\qwn| << 1$, and since $\rho_p({\vec x})$ is generally well 
determined from
parity conserving electron scattering, $\alr$ is essentially a direct \lq\lq 
meter" of the
Fourier transform of $\rho_n({\vec x})$. At low momentum-transfer ($qR_{N,P}<< 
1$) this
expression simplifies:
\begin{equation}
\label{theo:rhonb}
-\left[{4\sqrt{2}\pi\alpha\over G_F|q^2|}\right]\alr\approx {N\over 
Z}\left[1+{q^2\over 6}\left(
{R_P^2\over Z}-{R_N^2\over N}\right)\right]
\end{equation}
so that a determination of $R_N$ is, in principle, attainable from 
$\alr$\footnote{In a realistic
analysis of $\alr$ for heavy nuclei, the effects of electron wave distortion 
must be included
in the analysis of $\alr$. For a recent distorted wave calculation, see Ref. 
\cite{Hor98}}. 

In a realistic experiment PVES experiment, one does not have $qR_{N,P} << 1$; 
larger values of
$q$ are needed to obtain the requisite precision for reasonable running times 
\cite{Mus94,Don89}.
In Ref. \cite{Don89}, it was shown that a 1\% determination of $\rho_n({\vec 
x})$ for $^{208}$Pb
is experimentally feasible for $q\sim 0.5\ {\hbox{fm}}^{-1}$ with reasonable 
running times. An
experiment with barium is particularly attractive. If the barium isotopes are 
used in future
APV measurements as anticipated by the Seattle group, then a determination of 
$\rho_n({\vec
x})$ for even one isotope could reduce the degree of theoretical uncertainty for 
neutron
distributions along the barium isotope chain. Moreover, the first excited state 
of $^{138}$Ba
occurs at 1.44 MeV. The energy resolution therefore required to guarantee 
elastic scattering
from this nucleus is well within the capabilities of the Jefferson Lab.

\medskip
\noindent{\bf B. Hadronic Form Factors}
\medskip

From the form of Eq. (\ref{obs:alr}), it is clear that a precise determination 
of $\qpv$ from
$\alr$ requires sufficiently precise knowledge of the form factor term, $F(q)$. 
Since a
measurement can never be performed at kinematics for which $F(q)=0$, namely, 
$q^2=0$, it will
always generate a non-zero contribution to a PVES determination of $\qpv$. As 
discussed elsewhere
in this book, the
$F(q)$ term is presently under study at a variety of accelerators, with the hope 
of extracting
information on the strange quark matrix element $\bra{N(p')}{\bar s}\gamma_\mu 
s\ket{N(p)}$. The
latter is parameterized by two form factors, $\GES$ and $\GEp$. The other form 
factors which
enter $F(q)$ are known with much greater certainty than are the strange quark 
form factors. An
extraction of 
$\qpv$ from $F(q)$ requires at least one forward angle measurement 
\cite{Mus94,Mus92}. The
kinematics must be chosen so as to minimize the importance of $F(q)$ relative to 
$\qpv$ while
keeping the statistical uncertainty in the asymmetry sufficiently small. These 
competing kinematic
requirements -- along with the desired uncertainty in $\qpv$ -- dictate the 
maximum uncertainty
in $F(q)$ which can be tolerated. Since $\alr(^1H)$ generally manifests the 
greatest sensitivity
to new physics, I illustrate the form factor considerations for PV $ep$ 
scattering.

Since $Q_\sst{EM}^p=1$, the $ep$ asymmetry has the form
\begin{equation}
\label{theo:hadffa}
\alr=a_0\tau\ \left[\qwp+F^p(q)\right]\ \ \ ,
\end{equation}
where $a_0\approx 3.1\times 10^{-4}$ and $\tau=|q^2|/4\mns$. The form factor 
contribution
is given at tree level in the Standard Model by
\begin{equation}
\label{theo:hadffb}
F^p(\tau)=-\left[\GEp(\GEn+\GES)+\tau\GMp(\GMn+\GMS)\right]/\left[(\GEp)^2+\tau(
\GMp)^2\right]
\ \ \ ,
\end{equation}
where $G^{p,n}_\sst{E,M}$ denote the proton or neutron Sachs electric or 
magnetic form factors.
Since $Q_\sst{EM}^n=0$ and since the proton carries no net strangeness, both 
$\GEn$ and
$\GES$ must vanish at $\tau=0$. Consequently, one may write $F^p(\tau)$ as
\begin{equation}
\label{theo:hadffc}
F^p(\tau)=\tau B(\tau)\ \ \ .
\end{equation}
The function $B(\tau)$ will carry a non-trivial $\tau$-dependence if the 
$\tau$-dependence of
the strange-quark form factors differs from the  behavior observed for the 
nucleon
EM form factors. As noted in Refs. \cite{Mus94,Mus92}, however, any 
determination of $\qwp$ us be
made at such low-$\tau$ that only $B(\tau=0)$ enters the analysis, where 
\begin{equation}
\label{theo:hadffd}
b\equiv B(0)=(1-\mu_p)\mu_n-\mu_p\mu_s-\rhostr
\end{equation}
 and where
\begin{eqnarray}
\rhostr&=&{d\GES\over d\tau}\vert_{\tau=0}\\
\mu_s&=&\GMS(0)\ \ \ .
\end{eqnarray}
The low-$\tau$ form of the asymmetry  may be written as
\begin{equation}
\label{theo:hadffe}
{\alr\over a_0\tau}=\qwp+b\tau+{\cal O}(\tau^2)\ \ \ .
\end{equation}
 Extraction of $\qwp$ in this kinematic regime requires that the the uncertainty 
in
$b$ be minimized.

To set a scale for uncertainty in $b$, consider a 10\% determination of $\qpv$, 
which is
roughly comparable to the present $1.3\% $ determination of $\qpv({\hbox{Cs}})$. 

Consider also
possible  measurements at scattering angles $\theta=10^\circ$ and 
$\theta=5^\circ$.\footnote{These
choices correspond roughly to present ($\theta=12^\circ$) or prospective 
($\theta=6^\circ$)
Jefferson Lab capabilities.}  
Roughly one month of running could yield a 10\% determination of $\qwp$, 
provided one
maintains $\tau\geq 0.011$ ($\theta=10^\circ$) or $\tau\geq 0.003$ 
($\theta=5^\circ$). From
Eq. (\ref{theo:hadffe}) we infer the corresponding maximum uncertainty in $b$: 
$\delta b\leq 0.7$
for
$\theta=10^\circ$ and $\delta b\leq 2.5$. These numbers scale with the precision 
desired
for $\qwp$. 

The present program of PVES measurements at Jefferson Lab and the MAMI facility 
at Mainz
suggests that a determination of $b$ at this level is experimentally feasible. 
The HAPPEX
collaboration has reported a 15\% determination of the $ep$ asymmetry at 
$\tau=0.14$
\cite{happex}, which translates into an uncertainty in $B(\tau)$ of 0.3.  The 
translation of these results into uncertainties in $b$ requires some
care, and perhaps an additional measurement. If $B(\tau)$ varies gently with 
$\tau$ then one may
infer small uncertainties in $b$ from those in $B(\tau)$. Since a gentle 
$\tau$-dependence
cannot be assumed {\em a priori}, additional measurements to constrain the 
non-leading
$\tau$-dependence of $B(\tau)$ may be necessary. Nonetheless, the experimental 
needed to
sufficiently constrain $b$ appears possible. 

\medskip
\noindent{\bf C. Dispersion Corrections}
\medskip
\begin{figure}
\begin{center}
\epsfbox{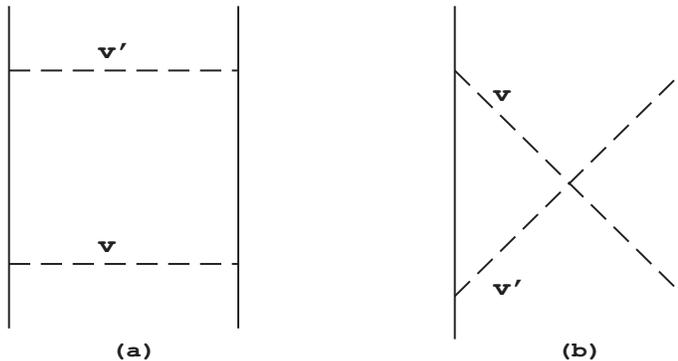}
\caption{\label{Fig 4} Two vector boson exchange dispersion corrections. Here
$V$ and $V'$ denote $\gamma$, $Z^0$, or $W^\pm$.}
\end{center}
\end{figure}

The discussion in this chapter has implicitly relied upon a one vector boson 
exchange (OVBE)
approximation of the electroweak amplitudes contributing to various processes. A 
realistic
analysis of precision observables must take into account contributions beyond 
the OVBE
amplitude. In the case of electron scattering, these contributions are generally 
divided into two
classes: Coulomb distortion of plane wave electron wavefunctions and dispersion 
corrections. The
former can be treated accurately for electron scattering using distorted wave 
methods
\cite{Hor98}. The dispersion correction, however, has proven less tractable. 

The leading dispersion correction (DC) arises from diagrams of Fig. 4, where the
intermediate state nucleus or hadron lives in any one of its excited states. 
More generally,
box diagrams like those of Fig. 4 can be treated exactly for scattering of 
electrons
from point like hadrons. When at least one of the exchanged bosons is a photon, 
the amplitude
is prone to infrared enhancements.  For elastic PV scattering of an electron 
from a point-like
proton, for example, the $Z-\gamma$  amplitude contains infrared enhancement 
factors such as
$\ln|s|/\mzs$, where $s$ is the $ep$ c.m. energy. Such factors can enhance the 
scale of the
amplitude by as much as an order of magnitude over the nominal ${\cal 
O}(\alpha)$ scale.
Consequently,  one might expect box graph amplitudes which depend on details of 
hadronic or
nuclear structure to be a potential source of theoretical error in the analysis 
of precision
electroweak observables.

Data on the electromagnetic dispersion correction for $ep$ scattering 
is in general agreement with the scale predicted by  theoretical calculations.
The situation regarding electron scattering from nuclei, however, is less 
satisfying.
Recent data $^{12}$C$(e,e')$ taken at MIT-Bates and NIHKEF, however, disagree 
dramatically
with nearly all published calculations (for a more detailed discussion and 
references,
see Ref. \cite{Mus93}). An experimental determination of any electroweak DC 
($V=\gamma$,
$V'=W^\pm,Z^0$) is unlikely, and reliance on theory to compute this correction 
is unavoidable.
As shown in Ref. \cite{MRM98}, the corresponding theoretical uncertainty is far 
less problematic
for a determination of $\qpv$ from PVES than for the extraction of information 
on the strange
quark form factors. The arguments leading to this conclusion are instructive, 
and I repeat them
here.

To this end, it is convenient to write the $(V,V')$ DC as a correction  
$R_{VV'}$ to the
tree level EM and PV neutral current ampltitudes:
\begin{eqnarray}
\label{theo:dc1}
M_\sst{EM}&=&M_\sst{EM}^\sst{TREE}[1+R_{\gamma\gamma}+\cdots]\\
\label{theo:dc2}
M_\sst{NC}^\sst{PV}&=&M_\sst{NC}^\sst{PV,\ TREE}[1+R_{VV'}+\cdots]\ \ \ ,
\end{eqnarray}
where $\cdots$ denotes other higher order corrections to the tree level 
amplitude. Because
$M_\sst{EM}^\sst{TREE}\propto 1/q^2$ while the $\gamma\gamma$ amplitude contains 
no pole
at $q^2=0$, $R_{\gamma\gamma}$ has the general structure
\begin{equation}
\label{theo:dc3}
R_{\gamma\gamma}(q^2)=q^2\ {\tilde R}_{\gamma\gamma}(q^2)
\end{equation}
where ${\tilde R}_{\gamma\gamma}(q^2)$ describes the $q^2$ dependence of the 
$\gamma\gamma$
amplitude and is finite at $q^2=0$. Since the tree level NC amplitude contains 
no pole at $q^2=0$,
however, the PV DC's do not vanish at $q^2=0$. Using Eq. (\ref{theo:dc3}) and 
expanding the PV
corrections in powers of
$q^2$ yields 
\begin{equation}
\label{theo:dc4}
{\alr\over a_0\tau}= \qpv\left[1+R_{WW}(0)+R_{ZZ}(0)+R_{Z\gamma}(0)\right]
+ {\tilde F}(q)
\end{equation}
where $F(q)$ in Eq. (\ref{obs:alr}) is replaced by an effective form factor 
${\tilde F}(q)$: 
\begin{equation}
\label{theo:dc5}
{\tilde F}(q)=F(q)+q^2\left[
R^{\prime}_{WW}(0)+R^{\prime}_{ZZ}(0)+R^{\prime}_{Z\gamma}(0)-{\tilde 
R}_{\gamma\gamma}(q^2)
+\cdots\right]\ \ \ .
\end{equation}
As before, $F(q)$ contains the dependence on hadronic and nuclear form factors 
discussed above.

From Eqs. (\ref{theo:dc3}-\ref{theo:dc5}) we can see that the entire 
$\gamma\gamma$ DC, as well as
the sub-leading
$q^2$-dependence of the $WW$, $ZZ$, and $Z\gamma$ DC's, contribute to $\alr$ as 
part of
an effective form factor term, ${\tilde F}(q)$. Since $F(q)\sim q^2$ for 
low-$|q^2|$ at
forward angles, the $\gamma\gamma$ DC contribution entering Eq. (\ref{obs:alr}) 
will be
exerimentally constrained along with $F(q)$ when the form factor term ${\tilde 
F}(q)$ is
kinematically separated from the weak charge term. Consequently, an extraction 
of $\qpv$ from
$\alr$ does not require  theoretical computations of the $\gamma\gamma$ DC or of 
the sub-leading
$q^2$-dependence of the other DC's. A determination of the strange-quark form 
factors, however,
will require such theoretical input. 

In order to extract constraints on possible new physics contributions to $\qpv$, 
one must compute $R_{WW}(0)$, $R_{ZZ}(0)$, and $R_{Z\gamma}(0)$. The theoretical
uncertainty associated with $R_{WW}(0)$ and $R_{ZZ}(0)$ is small, since box 
diagrams
involving the exchange are dominated by hadronic intermediate states having 
momenta
$p\sim M_\sst{W}$.  These contributions can be reliably treated perturbatively. 
The
$R_{Z\gamma}(0)$ correction, however, is infrared enhanced and displays a 
greater sensitivity to
the low-lying part of the nuclear and hadronic spectrum \cite{Mus90,Mar83}. 
Fortunately, the sum
of diagrams 4a and 4b conspire to suppress this contribution by 
$\gve=-1+4\sstw$. This feature
was first shown in Ref. \cite{Mar83} for the case of APV. Here, I summarize the 
argument as it
applies to both APV and scattering. 

The largest contributions to the loop integrals for diagrams 4a and 4b arise 
when
external particle masses and momenta are neglected relative to the loop momentum
$\ell_\mu$. The integrands from the two loop integrals sum to give  

\begin{eqnarray}
\label{theo:dc6}
{\bar u}[\gamma_\alpha{\not\ell}\gamma_\beta(\gve+\gae\gamma_5)
 &-&\gamma_\beta(\gve+\gae\gamma_5)
		{\not\ell}\gamma_\alpha] u\  T^{\alpha\beta}(\ell) D(\ell^2) \\
&=& 2i\ \epsilon_{\alpha\lambda\beta\mu}\ell^\lambda{\bar 
u}\gamma^\mu(\gve\gamma_5+\gae)u 
T^{\alpha\beta}(\ell) D(\ell^2)\ \ \ , \nonumber
\end{eqnarray}
where
\begin{equation}
\label{theo:dc7}
T^{\alpha\beta}(\ell) = \int d^4x\ \exp{i\ell\cdot x}\ 
\bra{0} T\left\{ J^\alpha_\sst{EM}(x) J^\beta_\sst{NC}(0)
\right\}\ket{0}\ \ \ ,
\end{equation}
$D(\ell^2)$ contains the electron and gauge boson propagators when external 
momenta and
masses are neglected relative to $\ell_\mu$, and $J^\alpha_\sst{EM}$ and 
$J^\beta_\sst{NC}$
are the hadronic electromagnetic and weak neutral currents, respectively. The 
terms in Eq.
(\ref{theo:dc6}) which transform like pseudoscalars are those containing the EM 
current and either
(a) both the axial currents ${\bar u}\gamma^\mu\gamma_5 u$ and 
$J_\sst{NC}^{\beta 5}$ or
(b) both the vector currents ${\bar u}\gamma^\mu u$ and $J_\sst{NC}^\beta$. The 
former
has the coefficient $Q_\sst{W}^e=-1+4\sstw$ and the latter has a the coefficient 
$\gae=1$.
The dependence of these terms on the spatial currents is given by ($\lambda=0$ 
in Eq.
(\ref{theo:dc7}))
\begin{eqnarray}
Q_\sst{W}^e\ {\hbox{term}}: &\sim& {\bar u}{\vec\gamma}\gamma_5 
u\cdot\left({\vec J}_\sst{EM}
	\times{\vec J}_\sst{NC}^5\right)\\
\gae\ {\hbox{term}}: &\sim& {\bar u}{\vec\gamma} u\cdot\left({\vec J}_\sst{EM}
	\times{\vec J}_\sst{NC}\right)\ \ \ .
\end{eqnarray}
The hadronic part of the $Q_\sst{W}$ term transforms as a polar vector, so that 
this
term contributes to the $A(e)\times V({\hbox{had}})$ amplitude. The hadronic 
part
of the $\gae$ terms, on the other hand, transfors as an axial vector, yielding
a contribution to the $V(e)\times A({\hbox{had}})$ amplitude. Hence, only the
$Q_\sst{W}^e$ term contributes to $\qpv$ term in the asymmetry. 

Since $Q_\sst{W}^e\sim -0.1$, the contribution $R_{Z\gamma}(0)$ in Eq. 
(\ref{theo:dc4}) is
suppressed with respect to the generic one-loop scale. Consequently, then,  the 
extraction
of new physics constraints from the first term in Eq. (\ref{theo:dc4}) will not 
be appreciably
affected by large theoretical uncertainties in the computation of 
$R_{Z\gamma}(0)$. In the case
of $\alr(^1{\hbox{H}})$, however, both $R_{Z\gamma}(0)$ and the tree-level 
amplitude are
proportional to $\qwe$, so that there exists no additional suppression of the 
former with
respect to the latter. Nevertheless, the theoretical computations of the 
$\gamma\gamma$
dispersion correction for parity-conserving $ep$ scattering appear to work 
reasonably well. 
Consequently, one would expect calculations of the $Z\gamma$ DC to be similarly 
reliable
for  PV $ep$ scattering. 

\section{Conclusions}
\label{sec:concl}

The realm of physics beyond the Standard Model offers rich possibilities for new 
discovery as
well as fertile ground for the development of new theoretical scenarios. As 
experiments begin to
probe this ground at TeV mass scales, low-energy PV can continue to play an 
important role in 
elucidating the larger framework in which the Standard Model must lie. Indeed, 
it is remarkable
that experiments performed with eV or a few GeV energies may have significant 
statements to make
about physics at the TeV scale. The motivation for pursuing the future of PV is 
undoubtedly high.

The future of low-energy PV presents serveral challenges to both 
experimentalists and theorists.
Improvements in new physics sensitivity will require progress on any one of a 
number of fronts.
If the 1\% cesium atomic theory uncertainty cannot be significantly improved 
upon, then the future
of APV may rest with measurements of isotope ratios. The interpretation of the 
${\cal R}_i$,
however, requires that nuclear theory and experiment achieve a more reliable 
determination of
$\rho_n(r)$ than presently exists. In this respect, PVES may prove to be an 
effective complement
to APV by providing a direct measurement of the neutron distribution. Precisely 
how such a
measurement would constrain the nuclear theory error entering the interpretation 
of ${\cal R}_i$
remains to be clearly delineated. From the standpoint of interpretability, PVES 
offers the
theoretically \lq\lq cleanest" probe of new physics. The theoretically most 
uncertain quantities
entering $\alr$ are either suppressed or can be measured by exploiting their 
kinematic dependence.
The prospects for improved new physics sensitivity is the highest with a forward 
angle
measurement on the proton, though the approved M\"oller experiment at SLAC and a 
possible
experiment with a $^4$He or $^{12}$C target would provide useful complements. It 
stands as a
challenge to experimentalists to achieve the necessary precision.

\section*{Acknowledgements}
%\newpage
%\begin{center}
%{\bf ACKNOWLEDGEMENTS}
%\end{center}

I wish to thank Profs. M.-A. Bouchiat and C. Bouchiat and the Ecole Normale for 
their
kind hospitality during the workshop. I also thank S.J. Pollock, W.J. Marciano, 
and D.
Budker for useful discussions and S.J. Puglia for assistance in preparing this 
manuscript.
This work was supported in part under U.S. Department of Department of Energy 
contract
\# DE-FG06-90ER40561 and a National Science Foundation Young Investigator Award. 

%\begin{center}
%{\bf REFERENCES}
%\end{center}

%\begin{figure}
%\epsfbox{fig1.eps}
%\caption{\label{Fig1} (a) Parity-violating electron-fermion amplitude generated 
%$Z^0$-exchange. (b) Effective, four-fermion electron-fermion PV interaction.}
%\end{figure}
%\medskip
%\begin{figure}
%\begin{center}
%\epsfbox{fig2.eps}
%\caption{\label{Fig2} Parity-violating semi-leptonic amplitude generated by 
%leptoquark
%(LQ) exchange.}
%\end{center}
%\end{figure}
%\begin{figure}
%\begin{center}
%\epsfbox{fig3.eps}
%\end{center}
%\caption{\label{Fig 3} Leptoquark (LQ) one-loop contributions to PV M\"oller 
%scattering.}
%\end{figure}
%\newpage
%\begin{figure}
%\begin{center}
%\epsfbox{fig4-2.eps}
%\caption{\label{Fig 4} Two vector boson exchange dispersion corrections. Here
%$V$ and $V'$ denote $\gamma$, $Z^0$, or $W^\pm$.}
%\end{center}
%\end{figure}

\end{document}